\documentclass[conference]{IEEEtran}
\IEEEoverridecommandlockouts
\usepackage{cite}
\usepackage{amsmath,amssymb,amsfonts}
\usepackage{breqn}
\usepackage{algorithmic}
\usepackage{graphicx}
\usepackage{subfig}
\usepackage{textcomp}
\usepackage{xcolor}
\def\BibTeX{{\rm B\kern-.05em{\sc i\kern-.025em b}\kern-.08em
    T\kern-.1667em\lower.7ex\hbox{E}\kern-.125emX}}

\DeclareMathOperator{\logm}{logm}

\DeclareMathOperator{\expm}{expm}

\begin{document}

\title{Comparison of Brain Networks based on Predictive Models of Connectivity
\thanks{This work was supported by EPSRC under Grant (EP/R026092/1). This work has received 'Best Student Paper Award AT IEEE BIBE 2019.}
}

\author{\IEEEauthorblockN{Fani Deligianni}
\IEEEauthorblockA{\textit{Hamlyn Center} \\
\textit{Imperial College}\\
London, UK \\
fani.deligianni@imperial.ac.uk}
\and
\IEEEauthorblockN{Jonathan D. Clayden}
\IEEEauthorblockA{\textit{Institute of Child Health} \\
\textit{UCL}\\
London, UK \\
j.clayden@ucl.ac.uk}
\and
\IEEEauthorblockN{Guang-Zhong Yang}
\IEEEauthorblockA{\textit{Hamlyn Centre} \\
\textit{Imperial College}\\
London, UK \\
g.z.yang@imperial.ac.uk}
}

\maketitle

\begin{abstract}

In this study we adopt predictive modelling to identify simultaneously commonalities and differences in multi-modal brain networks acquired within subjects. Typically, predictive modelling of functional connectomes from structural connectomes explores commonalities across multimodal imaging data. However, direct application of multivariate approaches such as sparse Canonical Correlation Analysis (sCCA) applies on the vectorised elements of functional connectivity across subjects and it does not guarantee that the predicted models of functional connectivity are Symmetric Positive Matrices (SPD). We suggest an elegant solution based on the transportation of the connectivity matrices on a Riemannian manifold, which notably improves the prediction performance of the model. 
Randomised lasso is used to alleviate the dependency of the sCCA on the lasso parameters and control the false positive rate.  
Subsequently, the binomial distribution is exploited to set a threshold statistic that reflects whether a connection is selected or rejected by chance. Finally, we estimate the sCCA loadings based on a de-noising approach that improves the estimation of the coefficients. 
We validate our approach based on resting-state fMRI and diffusion weighted MRI data. Quantitative validation of the prediction performance shows superior performance, whereas qualitative results of the identification process are promising.   

\end{abstract}

\begin{IEEEkeywords}
prediction, sparse CCA, functional connectomes, structural connectomes, model selection, identification, SPD, fMRI, Diffusion Weighted Images
\end{IEEEkeywords}

\section{Introduction}
Comparing neuroimaging data acquired from the same subjects is of paramount importance in understanding neuronal processes and extracting meaningful biomarkers in longitudinal and pharmacological studies \cite{rosa2015,ZALESKY2012,DADI2019,Milano2019}. However, both functional and structural brain connectomes comprise of thousands of connections, whereas the size of population does not exceed few tenths. Therefore, extracting meaningful relationships between networks is challenging and various methods have been developed to circumvent this problem \cite{bullmore2009,Watts1998,ZALESKY2012,zalesky:2010p3034,deligianni:2011ipmi,Deligiannipp2200}. 

One approach of comparing brain graphs is to use global properties of the network, such as small world index and clustering coefficient to describe its topological organisation and efficiency \cite{bullmore2009,Watts1998}. Global graph properties offer a robust way to compare brain networks, however, their interpretation is ambiguous and they are limited in that they cannot localise where the differences are \cite{neal_2017}. 

On the other hand, Zalesky et al. pioneered Network-Based Statistics (NBS) for extracting between-subjects differences across brain networks \cite{ZALESKY2012,zalesky:2010p3034}. Within-subject differences are modelled by adding explanatory variables for each subject. NBS performs univariate testing of the underlying hypothesis in each brain connection independently. To alleviate the problem of multiple comparisons, they use cluster-based thresholding and subsequently connected graph components are identified. This approach depends critically on a threshold statistic, which is chosen arbitrarily from the user. The hypothesis test is based on the component size, assuming that differences between groups of connectomes reflect systematic large network interactions.         

In this study, we suggest sparse predictive models of brain connectivity based on regularisation constraints to simultaneously extract within-subject similarities and differences between brain networks. Thus far, predictive models of multi-modal brain connectomes have been used to identify covariations across subjects \cite{honey:2009p3077,deligianni:2011ipmi,Deligiannipp2200,Deligianni14Fr,DADI2019}. Typically, they have been constructed based on multiple regression or Canonical Correlation Analysis (CCA) and regularisation.
Regularisation via $l_1$ constraints has attracted considerable interest in situations where the number of observations is much lower than the number of variables \cite{Witten:2009p5540,Varoquaux2013405,DADI2019}. In this case, sparse sets of associated variables would result in simultaneous multivariate dimensionality reduction and prediction. In fact one way, of establishing a relationship between structural brain connectomes $\mathbf{X}$ and functional brain connectomes $\mathbf{Y}$ is based on sparse Canonical Correlation Analysis (sCCA) \cite{Deligianni2016}.  
In sCCA, two sparse canonical vectors are extracted, subject to $l_1$ penalties, such that the projections of $\mathbf{X}$ and $\mathbf{Y}$ onto these vectors, respectively, are maximally linearly correlated \cite{Witten:2009p5540,witten2009}. 

However, sCCA is applied on vectorised versions of functional and structural connectivity across subjects. This affects functional connectivity, which is estimated as the precision matrix, the inverse of the covariance of the fMRI time-series, and it is restricted to a hypercone manifold of Symmetric Positive Definite (SPD) matrices. The problem of applying sCCA in the vectorised versions of connectivity elements is that there is no guarantee that the results from linear operations on the elements of an SPD matrix will also lie on an SPD manifold. This hinders our ability to estimate the geodesic distance between measured and predicted connectomes, and it affects prediction performance. Here, we suggest a solution to this problem by projecting the functional connectivity matrices $\mathbf{A}$ into the tangent space at a point $\mathbf{B} \in \mathcal{S}ym_p^+$, using the Log map \cite{pennec2006,Arsigny:2006p5775,varoquaux2010b}.

Prediction performance is used to set the sparsity level based on cross-validation. Subsequently, model identification is used to extract with some confidence the connections that are consistently selected or rejected.    
Towards this end, we use randomised Lasso and bootstrapping with replacement that resample the subjects space and provide an estimation of the probability a connection to be selected \cite{meinshausen:2010r,Deligiannipp2200,Deligianni2016}. A binomial distribution is formed that describes the probability of a connection to be present given the sparsity of the matrix. In this way, we form a threshold statistic that highlights connections that are accepted or rejected based on the estimated sparsity level. 

The coefficient of each brain connection reflects the corresponding loadings of the sCCA. Recent work has shown that coefficients derived from discriminative models, such as CCA, are prone to correlated noise and their direct interpretation could be misleading, even when the solution is sparse \cite{Haufe201496}. 
Here, we also extend sCCA to incorporate a step that improves the estimation of sCCA loading in each bootstrap iteration\cite{Haufe201496}.

Summarising, the key methodological contributions of this paper is to develop predictive models of brain connectivity to extract both similarities and differences between multi-modal brain networks. This is achieved based on \textit{i)} an approach to project the covariance matrices in a Riemannian manifold and perform sCCA in the tangent space, \textit{iii)} Randomised Lasso is adopted to relax the dependencies of the regularisation parameters on the underlying parameters, while it resamples the data to assign a probability in each connection, \textit{iv)} the binomial distribution is used to automatically set a threshold statistic of the identification results \textit{v)} the coefficient loading estimation is simplified by taking into consideration that only one sCCA component is considered.   

We validate our method by examining inter-relationships between functional brain connectomes derived from resting-state functional Magnetic Resonance Imaging (fMRI) and structural brain connectomes derived from Diffusion Weighted Magnetic Resonance Images (DWI). Microstructural indices are derived based on two diffusion models fitted in each voxel of the DWIs, namely the traditional tensor model and the Neurite Orientation Dispersion and Density Imaging (NODDI) model. The tensor model results in estimating structural connectomes that reflect fractional anisotropy (FA)and mean diffusivity (MD), whereas the NODDI model results in structural brain connectomes that reflect intracellular volume fraction (ICVF), the orientation dispersion index (ODI), the isotropic compartment (ISO) and the $\kappa$ parameter \cite{Zhang20121000}. Prediction of functional from structural connectomes improves across all microstructural indices when we constrain the predicted precision matrices to symmetric positive definite (SPD) space. Furthermore, identification results reveal relationships between microstructural indices, such as FA and $\kappa$, FA and ODI as well as MD and ICVF.

\section{Methods}

\subsection{ Transport on a Riemannian manifold to constrain the prediction to SPD}
\label{Transport}

We project the functional connectivity matrices $\mathbf{A}$ into the tangent space at a point $\mathbf{B} \in \mathcal{S}ym_p^+$, using the Log map \cite{pennec2006,Arsigny:2006p5775,varoquaux2010b}:
\begin{equation}
     Log_{b}(\mathbf{A})=\mathbf{B}^{1/2} \logm(\mathbf{B}^{-1/2} \mathbf{A} \mathbf{B}^{-1/2}) \mathbf{B}^{1/2} 
   \label{eqn:proj}
\end{equation}

where $\logm$ denotes matrix logarithm and $Log_{B}(\mathbf{A})$ is the tangent vector at $\mathbf{B}$, assuming that $\mathbf{B}$ is close to $\mathbf{A}$. 
This projection allows us to operate on these elements in a vectorised form, and project the prediction back to the space of SPD matrices using the inverse mapping, which guarantees positive definiteness:

\begin{equation}
     Exp_{b}(\mathbf{A})=\mathbf{B}^{1/2} \expm(\mathbf{B}^{-1/2} \mathbf{A} \mathbf{B}^{-1/2}) \mathbf{B}^{1/2} 
   \label{eqn:projR}
\end{equation}
Here, $\expm$ denotes matrix exponential.

However, if we naively apply eq. \ref{eqn:proj} into a tangent space $\mathbf{B}_s$ for each subject independently \cite{Arsigny:2006p5775}, each subject's projected covariance matrix will lie in a different tangent space \cite{ng2014}. We will therefore not be able to compare them with each other within the sCCA framework, and this will be reflected in the prediction performance of the algorithm. Instead, under the assumption that the average covariance matrix $\mathbf{A}$ is close to all subjects' individual matrices $\mathbf{A}_s$, we can use it to project all subject-specific covariance matrices into a common tangent space. 

We adapt this approach in a cross-validation of the sCCA method. Therefore, in each cross-validation loop we average the functional connectivity matrices (left-out subject not included), and we project each subject's covariance matrix into this tangent space. Subsequently, the prediction for the left-out subject is projected back into the space of SPD matrices, based on eq. \ref{eqn:projR}. Prediction performance is estimated based on the geodesic distance between measured and estimated functional connectomes \cite{meinshausen:2010r,deligianni:2011ipmi}:
\begin{equation}
     d_{AI}(\mathbf{C},\mathbf{D}) = tr(logm \mathbf{C}^{-1/2} \mathbf{D} \mathbf{C}^{-1/2} )^2    
   \label{eqn:perf}
\end{equation}

\subsection{ Identification of the most relevant connections }
\label{Identification}

Identification of the most relevant connections is important to provide biological interpretation of the mapping of structural to functional connections. However, devising a statistically sound way to accept or reject the null hypothesis is challenging because of the complexity of the underlying inference problem.
To this end, we have modified the biconvex criterion in sCCA \cite{witten2009} based on the randomised Lasso principle \cite{meinshausen:2010r}. This takes the following form:

 \begin{dmath}
		 u \leftarrow \mathrm{arg max_{u}} (w_{\mathbf{x}} \cdot u)^{T} \mathbf{\breve{X}}^{T}\mathbf{Y}v \;
		 \; \mathrm{subject  \: to}: \: { \|u\|^{2}\leq1, \: \|u\|_{1}\leq c_{1}, \:w_{\mathbf{x}} \in \{1,0.5\} }
\end{dmath}		

\begin{dmath}
	v \leftarrow \mathrm{arg max_{v}} u^{T} \mathbf{\breve{X}}^{T}\mathbf{Y} (v \cdot w_{\mathbf{y}}) \; \; \mathrm{subject  \: to}:\: { \|v\|^{2}\leq1, \: \|v\|_{1}\leq c_{2}, \:w_{\mathbf{y}} \in \{1,0.5\}  } 
   \label{eqn:scca_rand}
\end{dmath}

$w_{\mathbf{x}}$ and $w_{\mathbf{y}}$ are the coefficient weights chosen randomly equal to .5 or 1, as recommended by \cite{meinshausen:2010r,Deligianni2016}.
Furthermore, we use bootstrap with resampling to extract the connections that are consistently selected. In each bootstrap iteration we set the number of canonical variates to  $K=1$, which simplifies interpretation. Therefore, we estimate a probability of selecting a connection as the number of times the connection has been selected over the number of total bootstrap iterations. 
Note that $c_1$ and $c_2$ are chosen initially based on a permutation strategy and they remain the same through out the bootstrap iterations and across all microstructural indices. Therefore, sparsity of the sCCA loadings also remains relatively constant, and reflects the probability of a connection by chance. 

\subsection{Sparsity-based Threshold Statistic}
\label{Binomial}
We use the binomial distribution to test the null hypothesis for each connection:\textit{i)} a connection is selected by chance, \textit{ii)} a connection is rejected by chance. In this way, we can detect both the connections that are similar across modalities as well the connections that differ significantly. The binomial distribution has parameters $n$ and $p$. $p$ reflects the probability of success in a sequence of $n$ independent experiments. Here, $n$ is the number of randomised lasso iterations, whereas $p$ is the probability of a connection to be selected randomly based on the sparsity of the connectome.   

\subsection{Coefficients Estimation}
\label{coeff}
 
Finally, we extend our approach to incorporate a step that improves the estimation of sCCA loading in each bootstrap iteration as per \cite{Haufe201496}. 
\cite{Haufe201496} showed that when the number of canonical correlation components is equal to one, $K=1$, the estimation of coefficients $\mathbf{W}$ can be simplified:
 \begin{equation}
  \mathbf{W}\propto\Sigma_{\mathbf{X}}u
 \label{eqn:coefs}
\end{equation}
where $\Sigma_{\mathbf{X}}$ is the covariance matrix of the structural connections across subjects for each bootstrap iteration and $u$ are the loadings along structural connections estimated with sCCA. We adopt this method so that we not only assign a probability for each connection to be selected, but also a coefficient.

\subsection{Imaging}
\label{Imaging}
Imaging data was acquired from 19 healthy volunteers (mean age $32.6\pm7.8$ years) using a Siemens Avanto 1.5~T clinical scanner. 
Structural data was obtained using a self-shielded gradient set with maximum gradient amplitude of 40~mT~m$^{-1}$ and a 32-channel head coil. 
Three shells of DW-MRI with $b=2400$~s~mm${}^{-2}$ (60 noncollinear gradient directions and six b0 images), $b=800$~s~mm${}^{-2}$ (30 noncollinear gradient directions and three b0 images) and $b=300$~s~mm${}^{-2}$ (9 noncollinear gradient directions and one b0 image) were acquired with a voxel matrix of 96$\times$96, 60 contiguous axial slices, each 2.5~mm thick, with 240$\times$240$\times$150~mm field of view (FOV), voxel size of 2.5$\times$2.5$\times$2.5~mm and TR/TE${}=8300$/98~ms.  
This imaging protocol was optimised for the acquisition of NODDI data on a 1.5~T scanner, and it requires approximately 16 minutes.

Resting-state fMRI data were acquired based on a T2*-weighted gradient-echo EPI sequence with 300 volumes, TR/TE${}=2160$/30~ms, 30 slices with thickness 3.0~mm (1~mm gap), effective voxel size 3.3$\times$3.3$\times$4.0~mm, flip angle 75$^\circ$, FOV 210$\times$210$\times$120 mm. High resolution T1-weighted whole-brain structural images were also obtained in all subjects with voxel size of 1.0$\times$1.0$\times$1.0~mm, TR/TE${}=11$/4.94~ms, flip angle 15$^\circ$, FOV 256$\times$256$\times$256 mm, voxel matrix 176$\times$216 and 256 contiguous slices. Ethical approval was obtained from the UCL Research Ethics Committee and informed consent obtained from all subjects. The raw data are available online \cite{datasetAll}.

\begin{figure*}[!ht]

\centering

\parbox{1.0\textwidth}{\centering \includegraphics[width=1.0\textwidth]{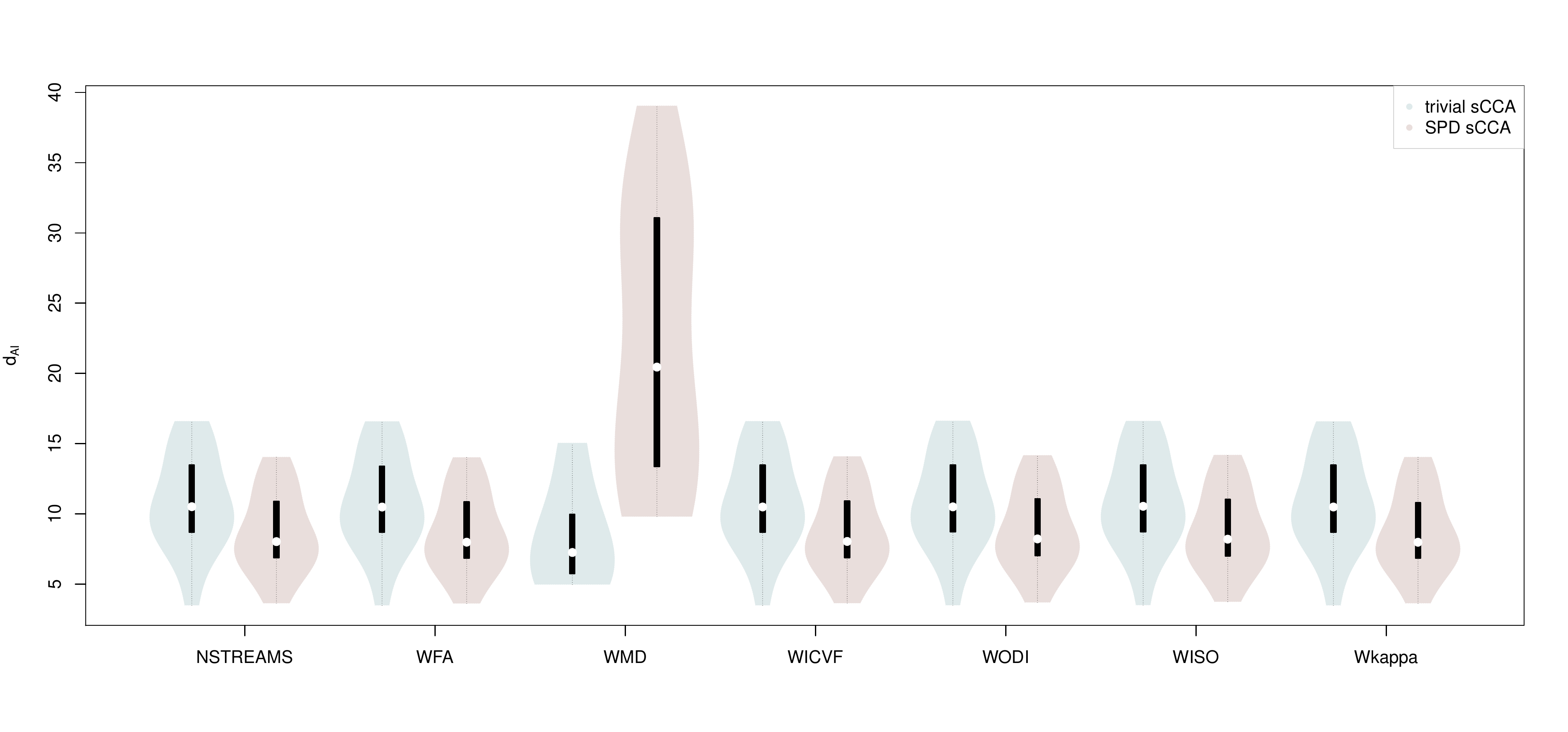}}

\textbf{\refstepcounter{figure}\label{fig:pred_perf} Figure \arabic{figure}.}{ Prediction performances of trivial sCCA and SPD sCCA for each of the microstructural indices. $d_{AI}$ is a distance metric between predicted and measured functional connectomes. The smaller the distance, the better the prediction.}
\end{figure*}

\subsubsection{Estimating functional and structural brain connectomes}
\label{connectomes}
We used TractoR for preprocessing of the diffusion weighted images \cite{Clayden:2011p5590,parker2014_PLOSONE}. Fractional anisotropy (FA) and mean diffusivity (MD) were estimated based on a tensor model fitted in each voxel. 
For the NODDI data we concatenated the three shells with $b$-values of 2400~s~mm${}^{-2}$, 800~s~mm${}^{-2}$ and 300~s~mm${}^{-2}$ in one shell, along with the b0 images acquired in each shell. Subsequently, we used TractoR, which wraps FSL, for eddy current correction along the concatenated volume of images. We used the NODDI Matlab toolbox to extract the intracellular volume fraction (ICVF), the orientation dispersion index (ODI), the isotropic compartment (ISO) and the $\kappa$ parameter \cite{Zhang20121000}. 
T1-weighted images were processed with Freesurfer to obtain 68 cortical grey matter (GM) regions \cite{Desikan2006968}. We propagated the anatomical labels from T1 space to both DWI space and fMRI space with non-rigid and affine registration, respectively \cite{Modat:2010p5093}.
We used FSL for basic preprocessing of the rs-fMRI data \cite{smith2004}. To construct corresponding functional networks, the fMRI signal is averaged across voxels within each GM ROI derived from the parcellation. 
The signal in white matter and cerebrospinal fluid is also averaged and, along with the six motion parameters provided by FSL, is linearly regressed out from the averaged time-series within each GM ROI.
We used the inverse covariance, normalised to unit diagonal, to characterise functional connectivity. This so-called precision matrix reflects partial correlation \cite{Smith:2013p5726}. 

A ball and two sticks multi-compartment fibre model was fitted to the shell of diffusion data acquired with $b=2400$~s~mm${}^{-2}$ with FSL \cite{behrens:2003p3099}.
Subsequently, we ran probabilistic tractography, implemented in TractoR, for each dataset. We seed 100 streamlines from each white matter voxel, and tracking terminates when both ends of a track reach a cortical target region.   

Structural brain connectomes are described as graphs, with nodes corresponding to ROIs and edges defined based on either the number of streamlines or as weighted averages of microstructural indices. In weighted averages, weights reflect the number of streamlines that pass through each voxel of the tract. Structural brain connectomes are derived based on the following microstructural indices: \textit{i)} As the number of fibres that connect the regions, divided by the average number of voxels within the two end-point ROIs (NSTREAMS).
\textit{ii)} As the weighted average of FA along the streamlines that connect the two regions (WFA).
\textit{iii)} As the weighted average of MD along the streamlines that connect the two regions (WMD). 
\textit{iv)} As the weighted average of ICVF along the streamlines that connect the two regions (WICVF). 
\textit{v)} As the weighted average of ODI along the streamlines that connect the two regions (WODI). 
\textit{vi)} As the weighted average of ISO along the streamlines that connect the two regions (WISO). 
\textit{vii)} As the weighted average of the $\kappa$ along the streamlines that connect the two regions (Wkappa).

\begin{figure*}[!h]
\centering
\subfloat[PCA loading]{\label{fig:pcaL} \includegraphics[trim=0.2cm 1.3cm 0.1cm 0.5cm, clip=true, width=0.45\textwidth]{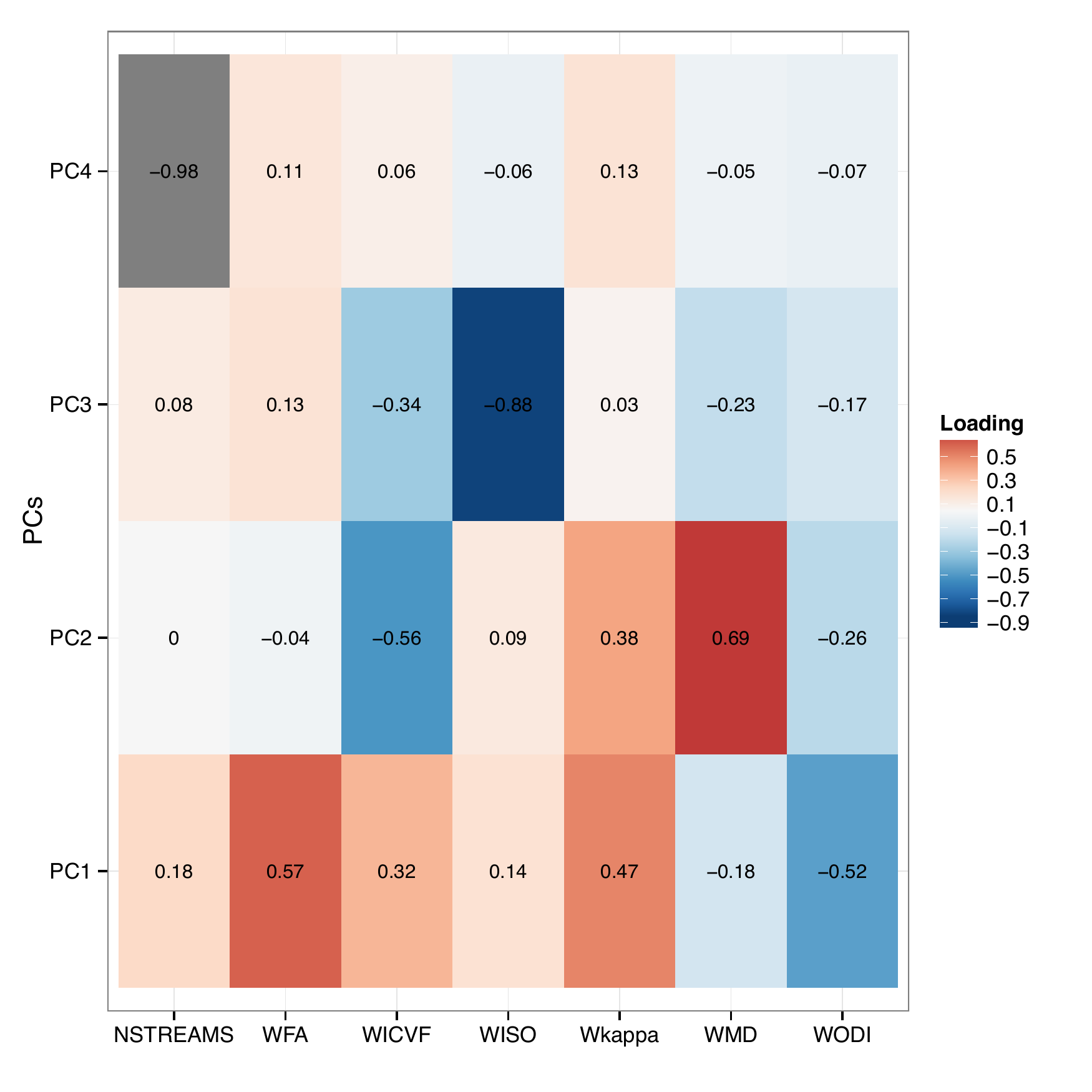} }
\centering
\subfloat[PCA Variances]{\label{fig:pcaV} \includegraphics[trim=1.1cm 1.6cm 1.0cm 2.0cm, clip=true, width=0.45\textwidth]{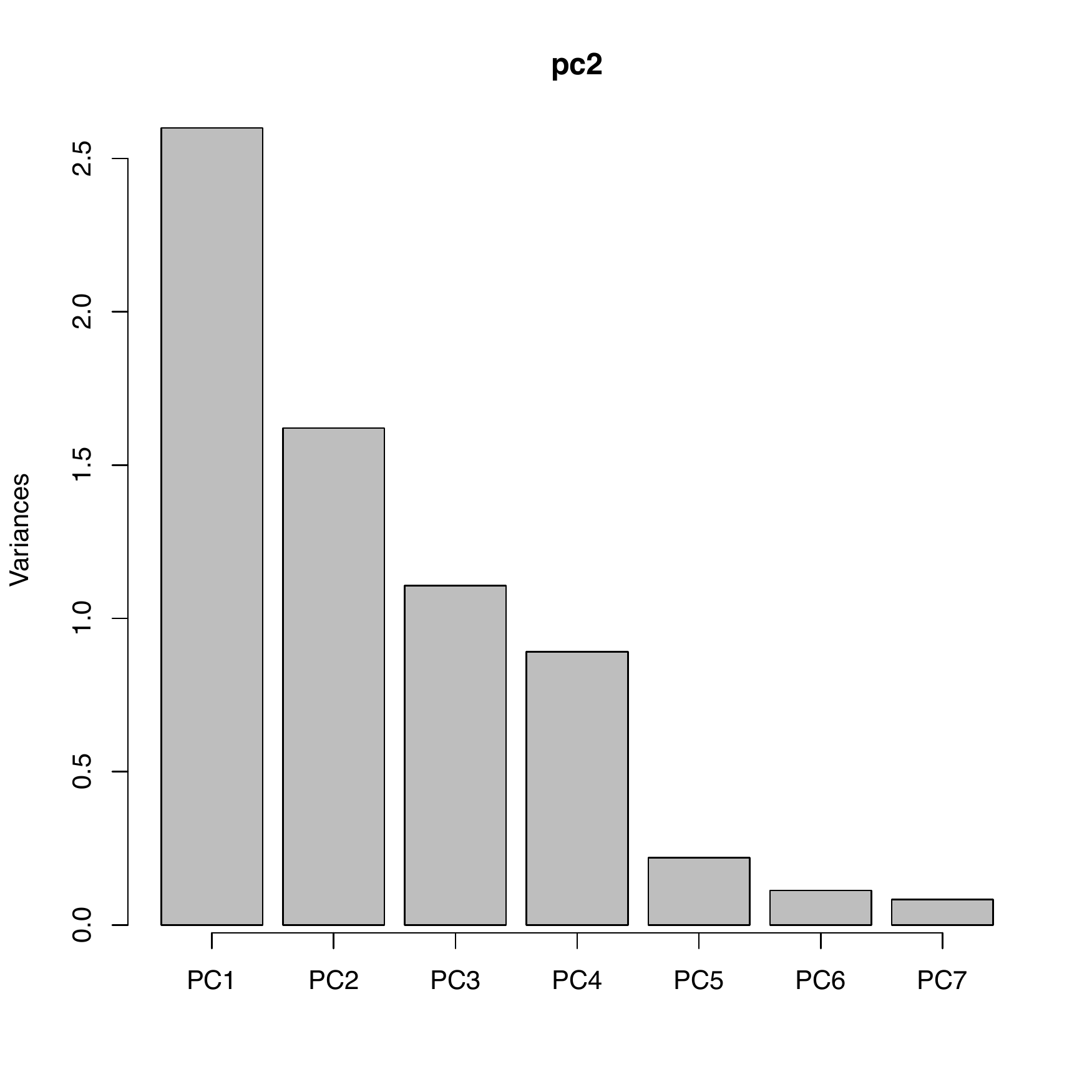} }

\hspace{5.0 mm}

\textbf{\refstepcounter{figure}\label{fig:pca} Figure \arabic{figure}.}{ PCA on the concatenated connectivity values along subjects.}
\end{figure*}

\section{Results}
\label{Results}

Figure \ref{fig:pred_perf} summarises the prediction performance of trivial sCCA (direct application of sCCA on the upper triangular part of connectivity matrices, which have been vectorised and columnwise concatenated across subjects) and SPD sCCA (functional connectomes have been projected onto a common tangent space, eq. \ref{eqn:proj}, before the application of sCCA), for each of the microstructural indices. SPD sCCA outperforms trivial sCCA in prediction performance and the pairwise differences are statistically significant based on a paired Wilcoxon rank test across all microstructural indices apart from WMD. In WMD the difference in prediction performance is not statistically significant because they are based only on three measurements, since trivial sCCA results in not SPD functional connectivity matrices. On the other hand, SPD sCCA provides 18 SPD predictions out of 19 cross-validation loops. 

To better understand how microstructural indices relate across subjects, we concatenated the connectivity values across all subjects into columnwise vectors for each index. Figure \ref{fig:pca} shows the loadings for the first four principal components (PCs). We observe that WFA is relatively correlated with Wkappa and anti-correlated with WODI, whereas WMD is anti-correlated to WICVF.

\begin{figure*}

\centering 
\vspace{1mm}
\parbox{0.01\textwidth}{\centering\hfill}
\parbox{0.2\textwidth}{\centering \scriptsize{\Large{all}} }
\parbox{0.2\textwidth}{\centering \scriptsize{\Large{rejected}} }
\parbox{0.2\textwidth}{\centering \scriptsize{\Large{remaining}} }
\parbox{0.2\textwidth}{\centering \scriptsize{\Large{accepted}} }
\parbox{0.05\textwidth}{\centering\hfill}

\centering 
\vspace{1mm}
\parbox{0.01\textwidth}{\centering \rotatebox[origin=c]{90}{\scriptsize{\large{NSTREAMS} }} } 
\parbox{0.2\textwidth}{\centering \includegraphics[trim=6.5cm 6.3cm 6.5cm 5.9cm, clip=true, width=0.2\textwidth]{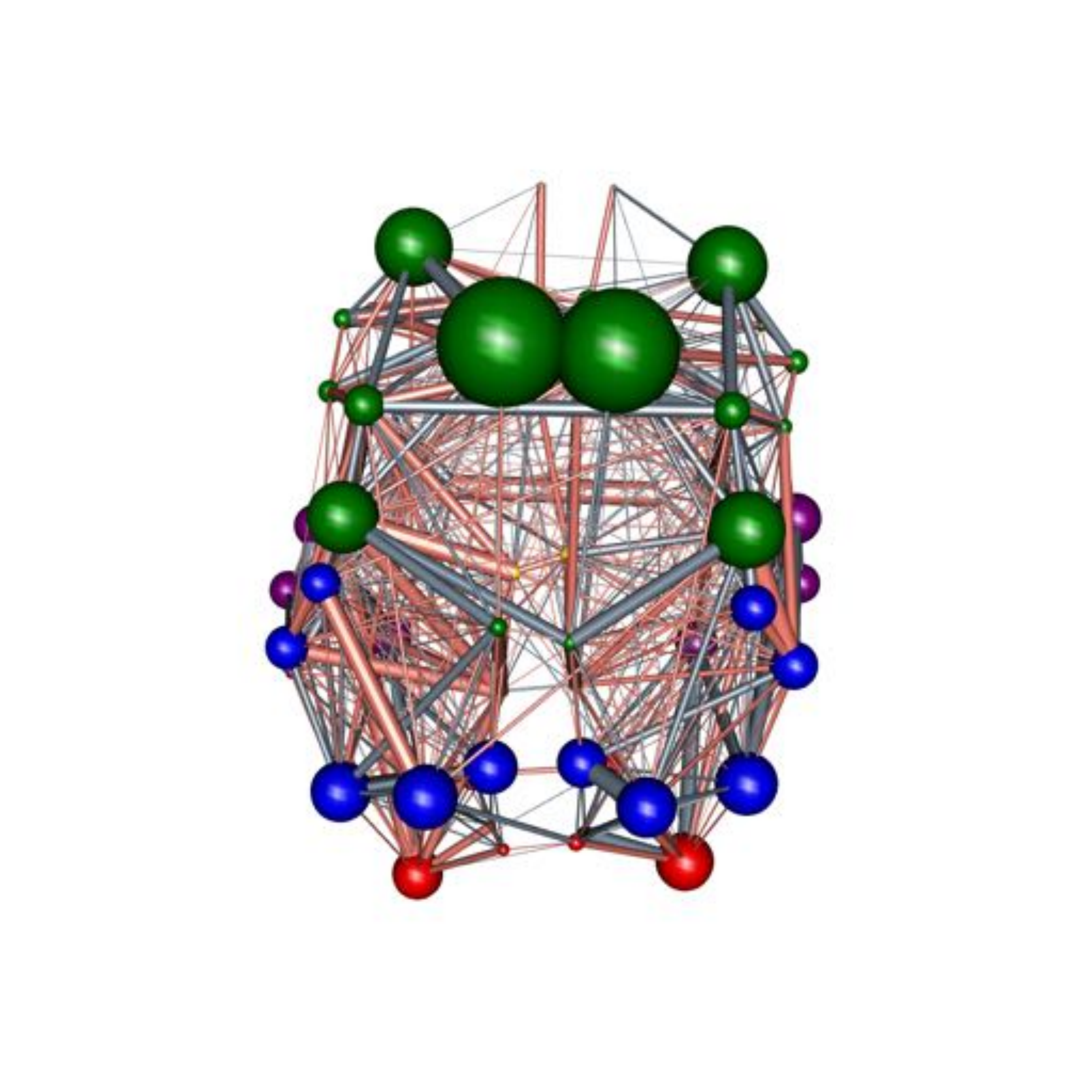}  }  
\parbox{0.2\textwidth}{\centering \includegraphics[trim=6.5cm 6.3cm 6.5cm 5.9cm, clip=true, width=0.2\textwidth]{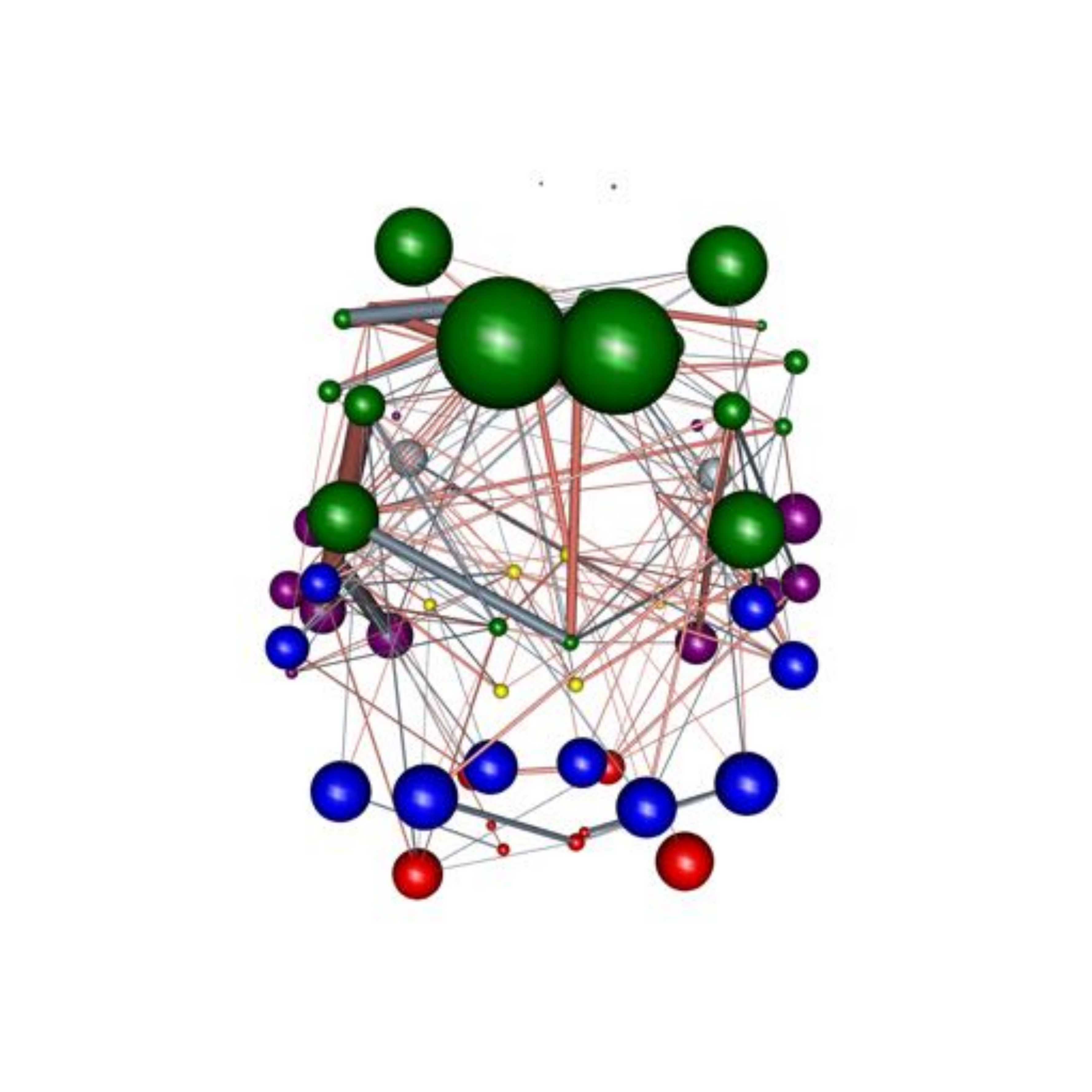}  }  
\parbox{0.2\textwidth}{\centering \includegraphics[trim=6.5cm 6.3cm 6.5cm 5.9cm, clip=true, width=0.2\textwidth]{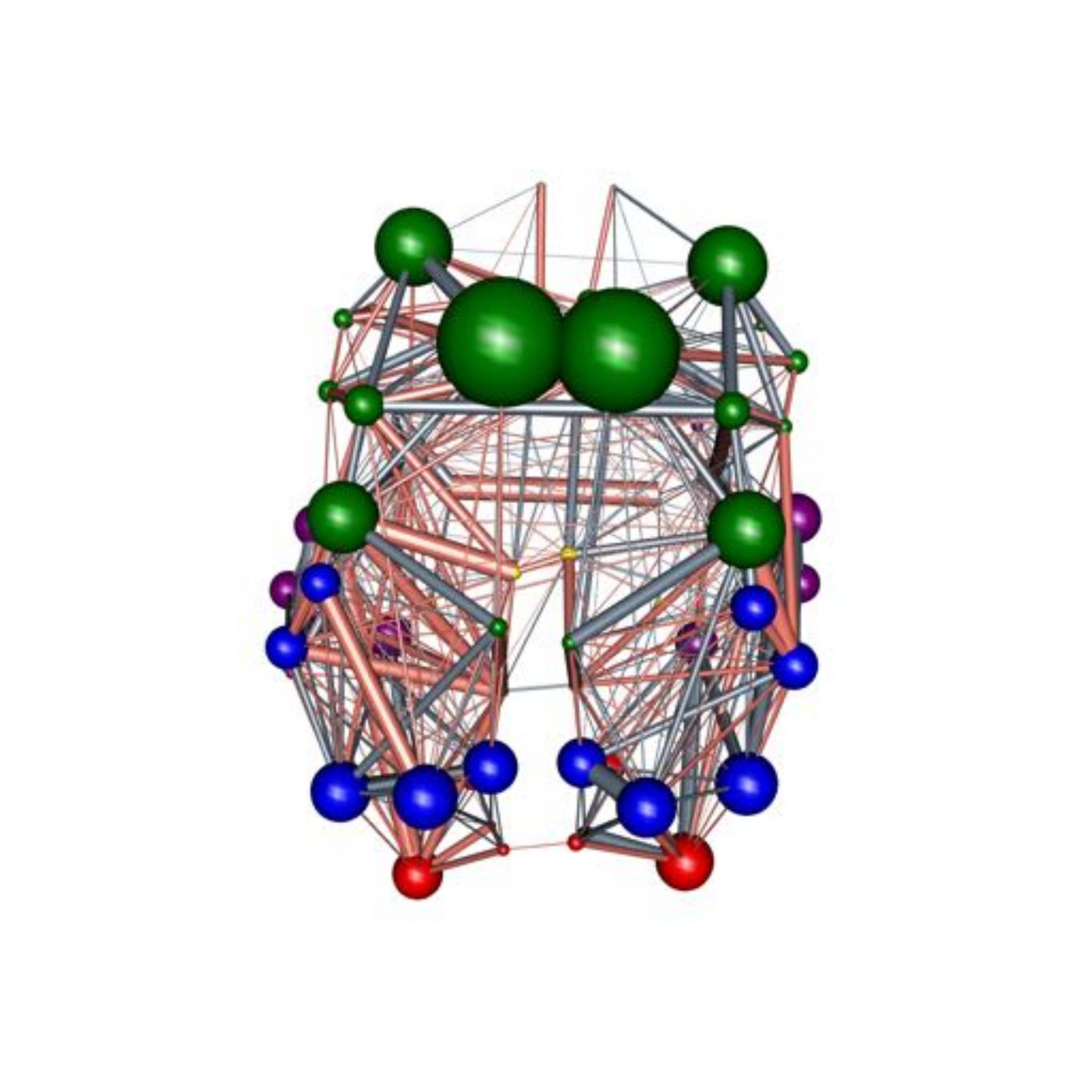}  }  
\parbox{0.2\textwidth}{\centering \includegraphics[trim=6.5cm 6.3cm 6.5cm 5.9cm, clip=true, width=0.2\textwidth]{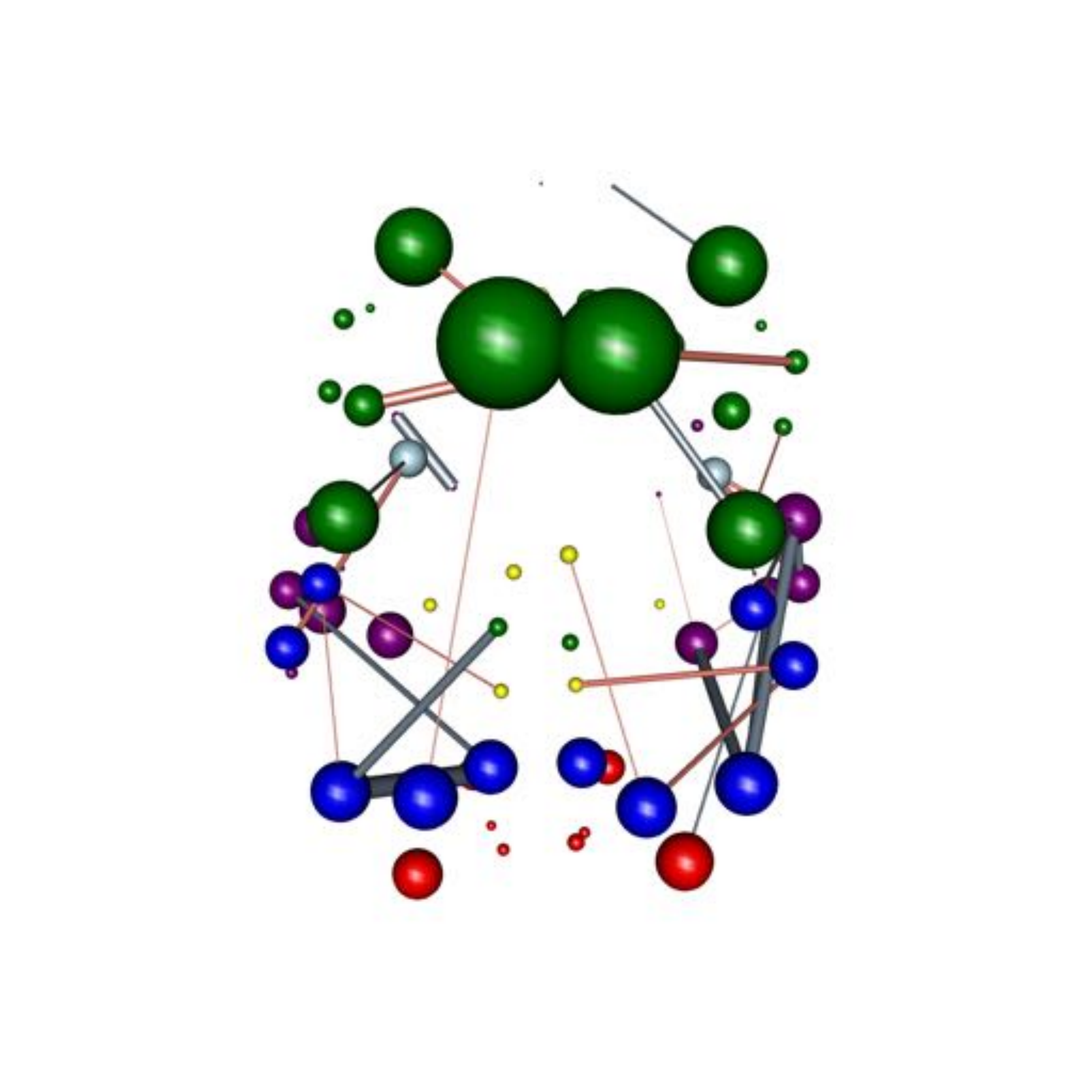}  }  
\parbox{0.05\textwidth}{\centering \includegraphics[width=0.04\textwidth]{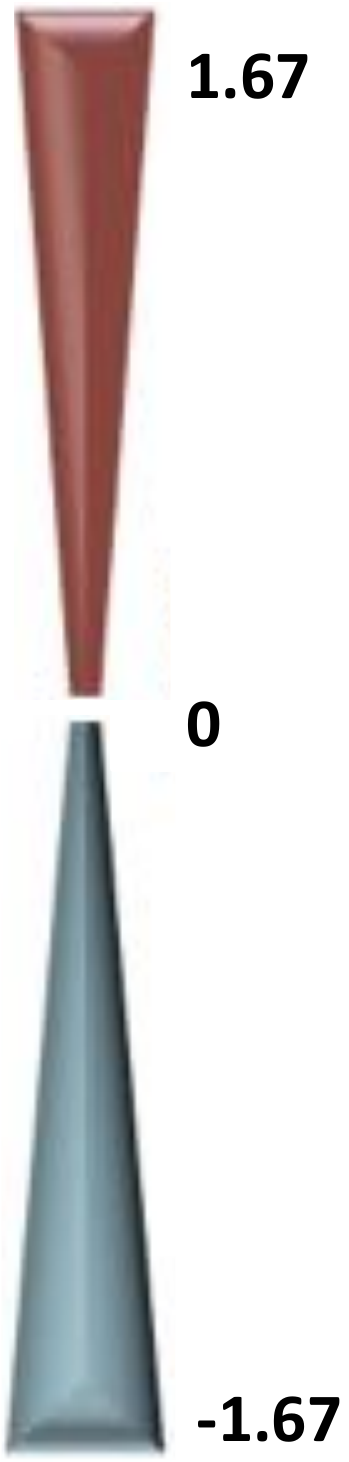}  } 

\centering 
\vspace{1mm}
\parbox{0.01\textwidth}{\centering \rotatebox[origin=c]{90}{\scriptsize{ \large{WFA} }} } 
\parbox{0.2\textwidth}{\centering \includegraphics[trim=6.5cm 6.3cm 6.5cm 5.9cm, clip=true, width=0.2\textwidth]{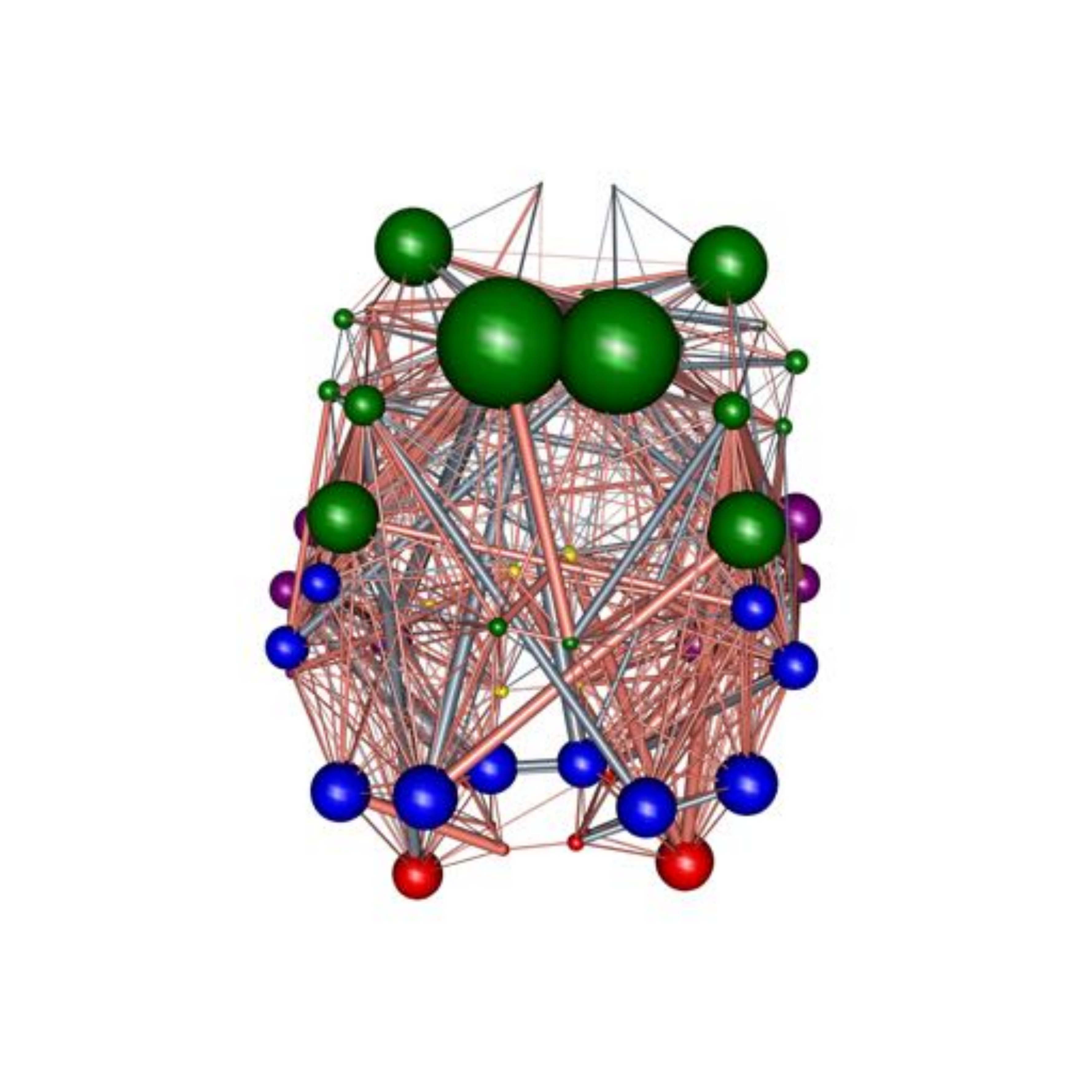}  }  
\parbox{0.2\textwidth}{\centering \includegraphics[trim=6.5cm 6.3cm 6.5cm 5.9cm, clip=true, width=0.2\textwidth]{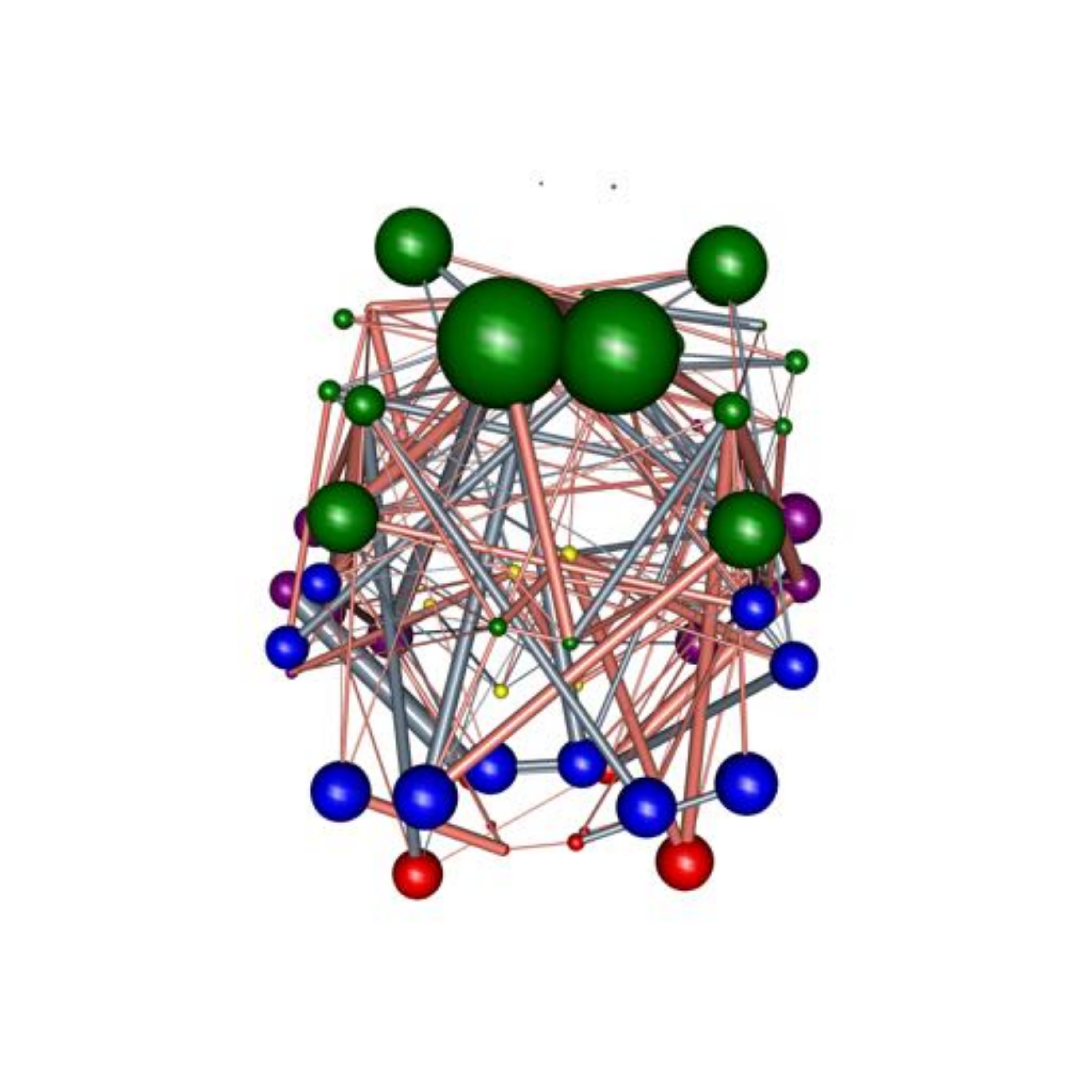}  }  
\parbox{0.2\textwidth}{\centering \includegraphics[trim=6.5cm 6.3cm 6.5cm 5.9cm, clip=true, width=0.2\textwidth]{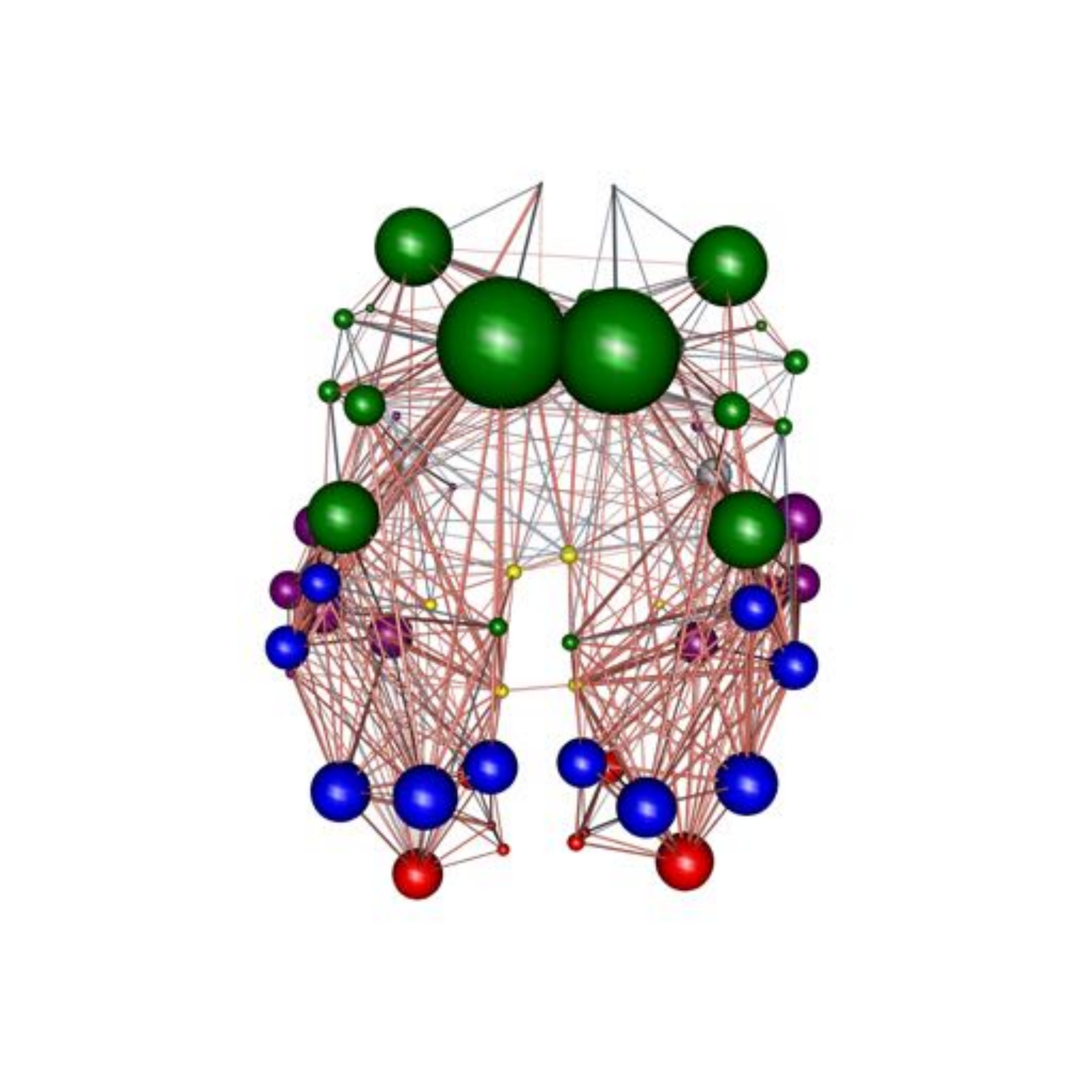}  }  
\parbox{0.2\textwidth}{\centering \includegraphics[trim=6.5cm 6.3cm 6.5cm 5.9cm, clip=true, width=0.2\textwidth]{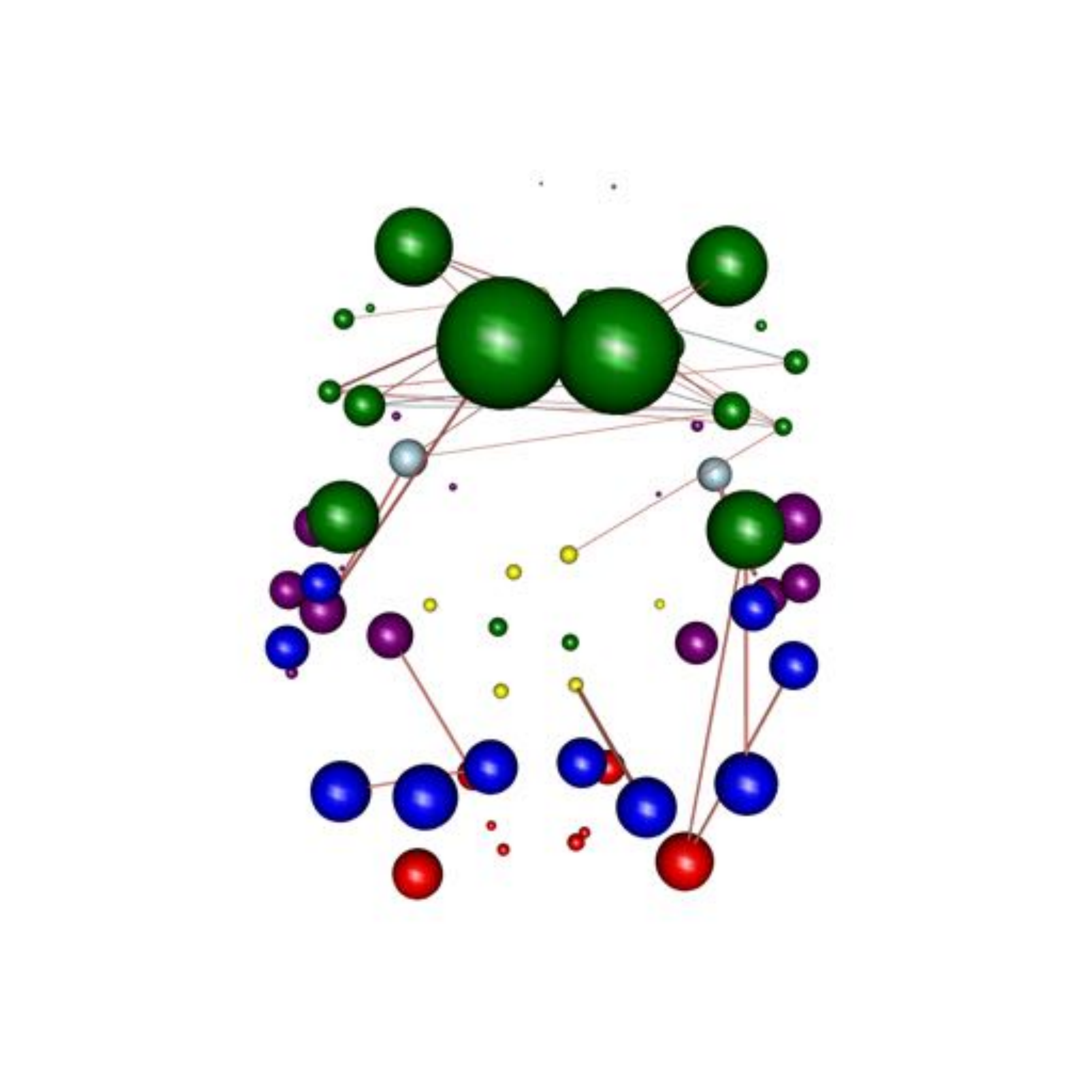}  }  
\parbox{0.05\textwidth}{\centering \includegraphics[width=0.04\textwidth]{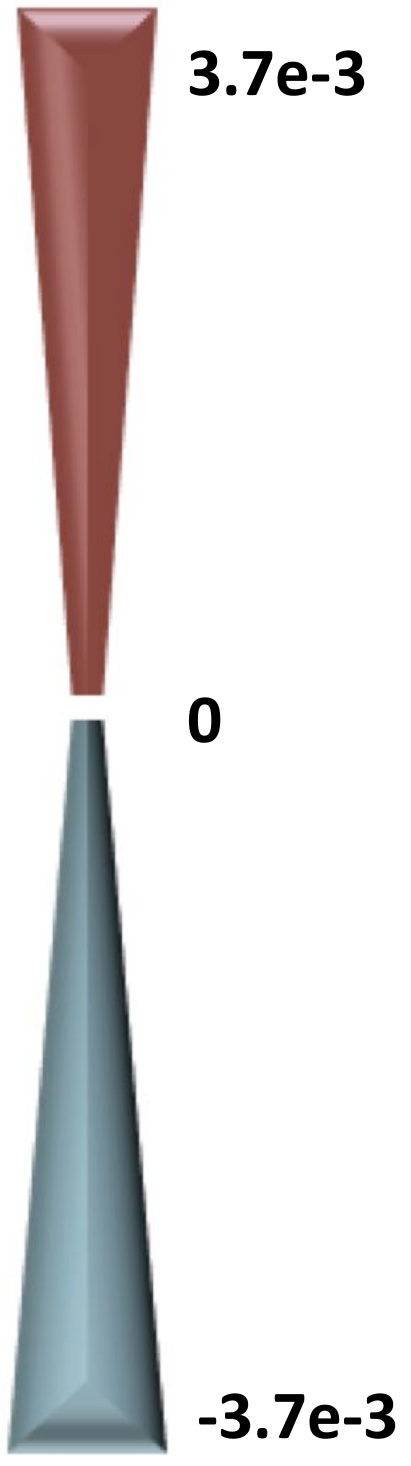}  } 

\centering 
\vspace{1mm}
\parbox{0.01\textwidth}{\centering \rotatebox[origin=c]{90}{\scriptsize{ \large{WMD} }} } 
\parbox{0.2\textwidth}{\centering \includegraphics[trim=6.5cm 6.3cm 6.5cm 5.9cm, clip=true, width=0.2\textwidth]{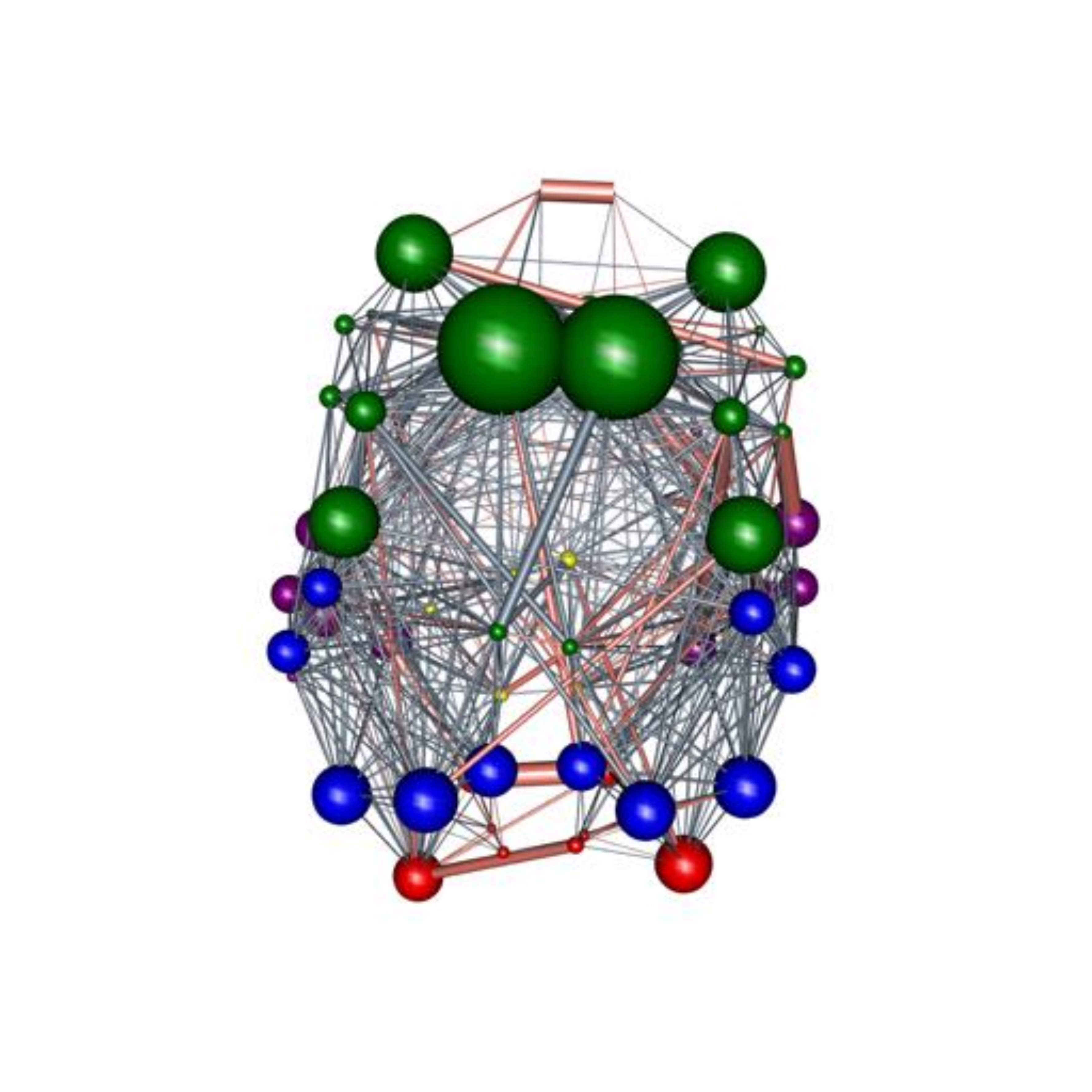}  }  
\parbox{0.2\textwidth}{\centering \includegraphics[trim=6.5cm 6.3cm 6.5cm 5.9cm, clip=true, width=0.2\textwidth]{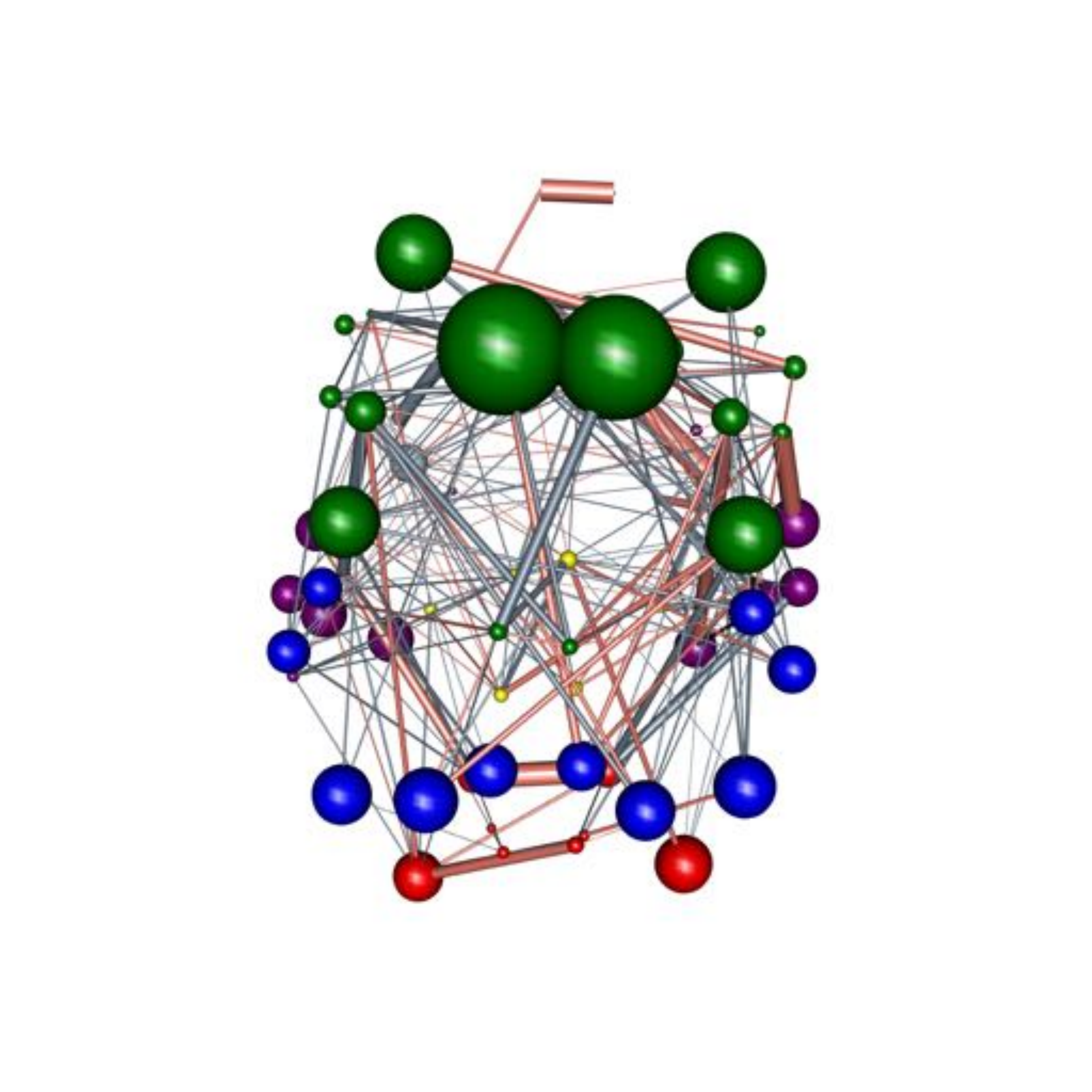}  }  
\parbox{0.2\textwidth}{\centering \includegraphics[trim=6.5cm 6.3cm 6.5cm 5.9cm, clip=true, width=0.2\textwidth]{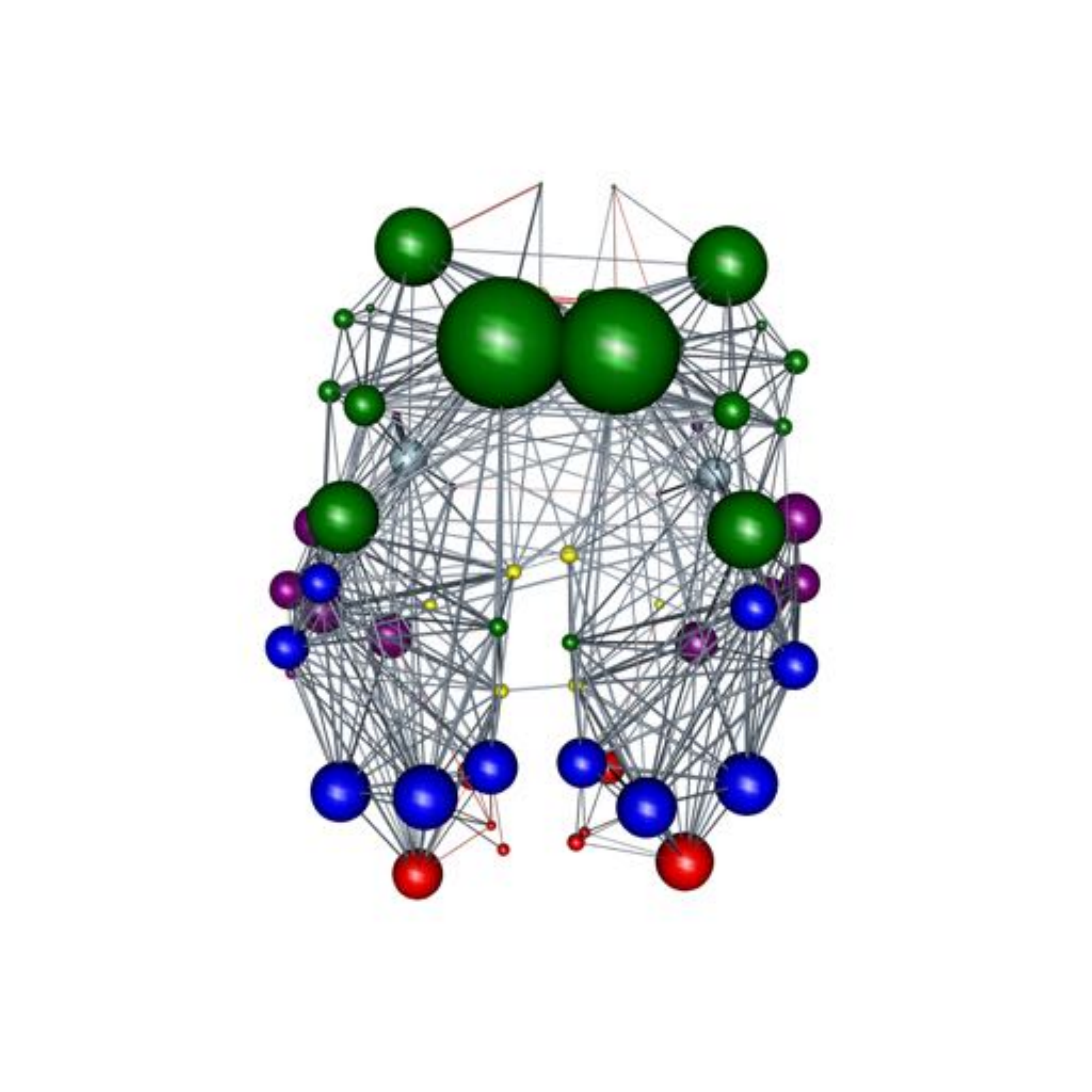}  }  
\parbox{0.2\textwidth}{\centering \includegraphics[trim=6.5cm 6.3cm 6.5cm 5.9cm, clip=true, width=0.2\textwidth]{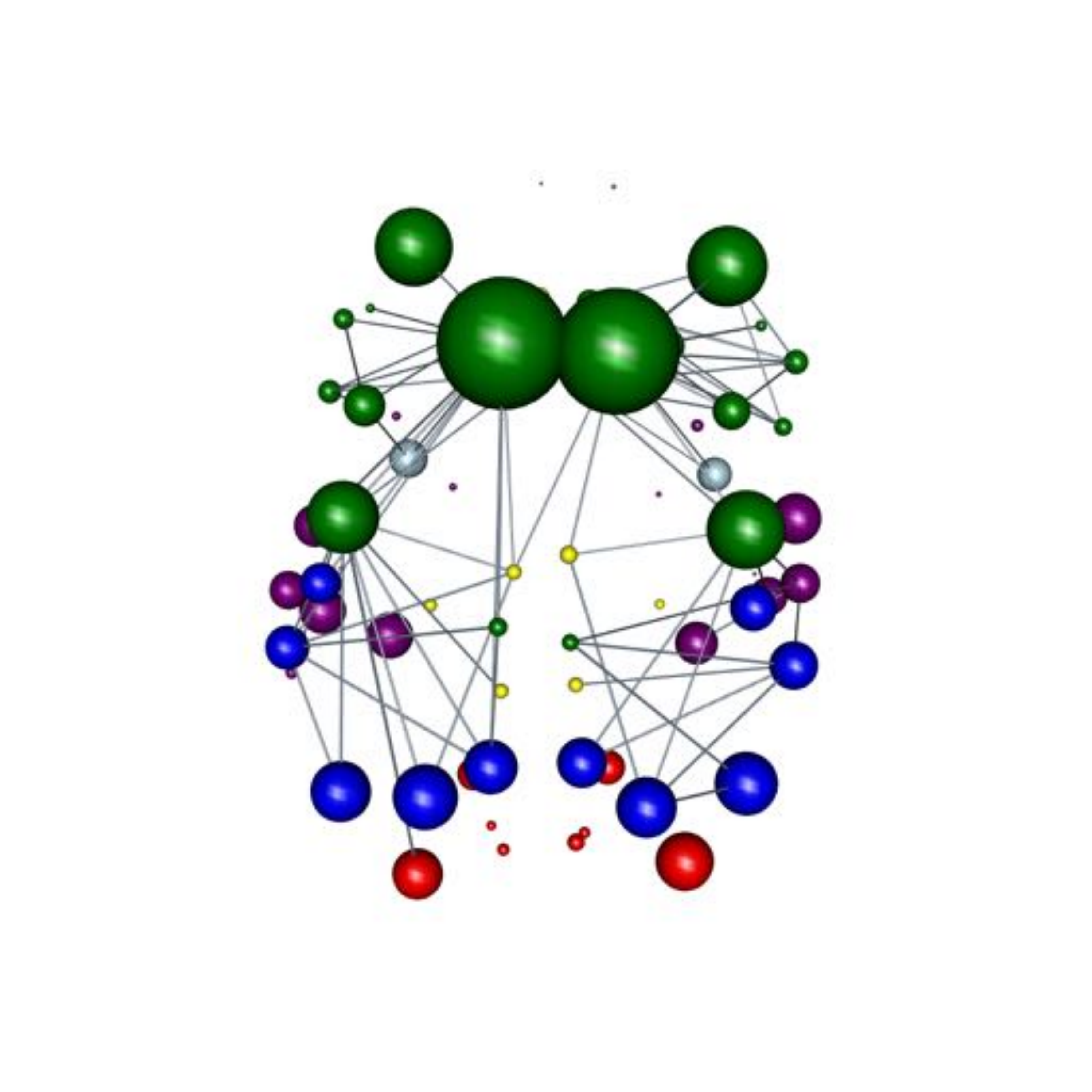}  }  
\parbox{0.05\textwidth}{\centering \includegraphics[width=0.04\textwidth]{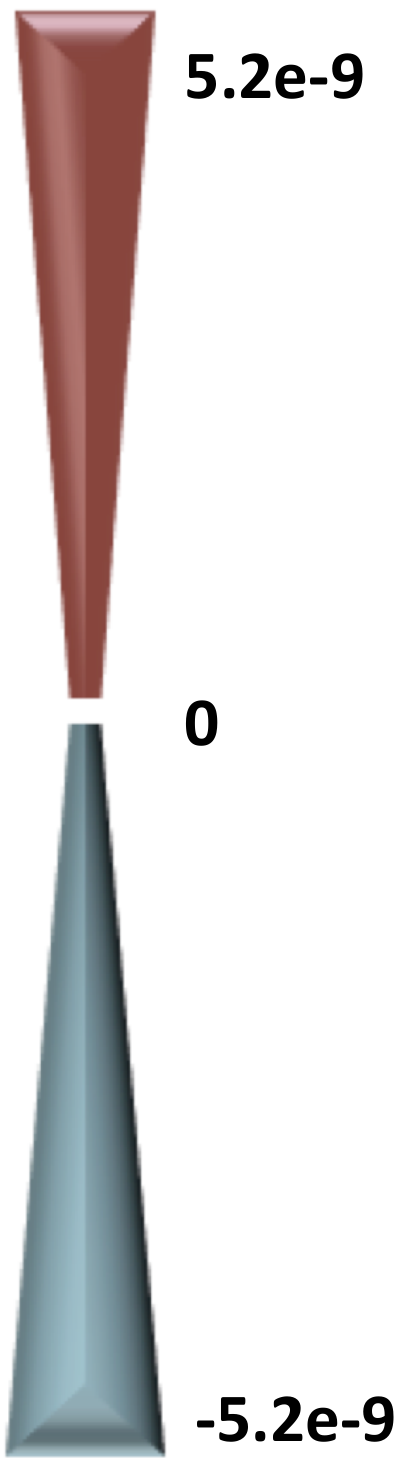}  } 

\centering 
\vspace{1mm}
\parbox{0.105\textwidth}{\centering \includegraphics[width=0.02\textwidth]{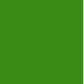} frontal}  
\parbox{0.105\textwidth}{\centering \includegraphics[width=0.02\textwidth]{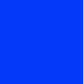}   parietal} 
\parbox{0.115\textwidth}{\centering \includegraphics[width=0.02\textwidth]{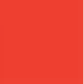}  occipital}
\parbox{0.125\textwidth}{\centering \includegraphics[width=0.02\textwidth]{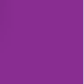}  temporal}
\parbox{0.105\textwidth}{\centering \includegraphics[width=0.02\textwidth]{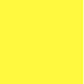} limbic}  
\parbox{0.105\textwidth}{\centering \includegraphics[width=0.02\textwidth]{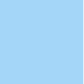}  insula} 

\hspace{0.1mm}

\textbf{\refstepcounter{figure}\label{fig:logmIdent} Figure \arabic{figure}.}{ Identification results for NSTREAMS and tensor based microstructural indices, namely  WFA, WMD. The first column shows the corrected coefficients for all connections with non-zero values across all subjects. The second column shows the connections that are rejected with a significant p-value ($<0.05$) based on a binomial distribution. The probability of a connection to be selected randomly is defined based on the sparsity of the connectomes and it is equal to 0.04. The third column shows the remaining connections. The forth column shows the connections that are selected significantly above chance (p-value$<0.05$) according to a binomial distribution.}
\end{figure*}

\begin{figure*}

\centering 
\vspace{1mm}
\parbox{0.01\textwidth}{\centering\hfill}
\parbox{0.2\textwidth}{\centering \scriptsize{ \Large{all} } }
\parbox{0.2\textwidth}{\centering \scriptsize{ \Large{rejected} } }
\parbox{0.2\textwidth}{\centering \scriptsize{ \Large{remaining} } }
\parbox{0.2\textwidth}{\centering \scriptsize{ \Large{accepted} } }
\parbox{0.05\textwidth}{\centering\hfill}

\centering 
\vspace{1mm}
\parbox{0.01\textwidth}{\centering \rotatebox[origin=c]{90}{\scriptsize{ \large{WICVF} }} } 
\parbox{0.2\textwidth}{\centering \includegraphics[trim=6.5cm 6.3cm 6.5cm 5.9cm, clip=true, width=0.2\textwidth]{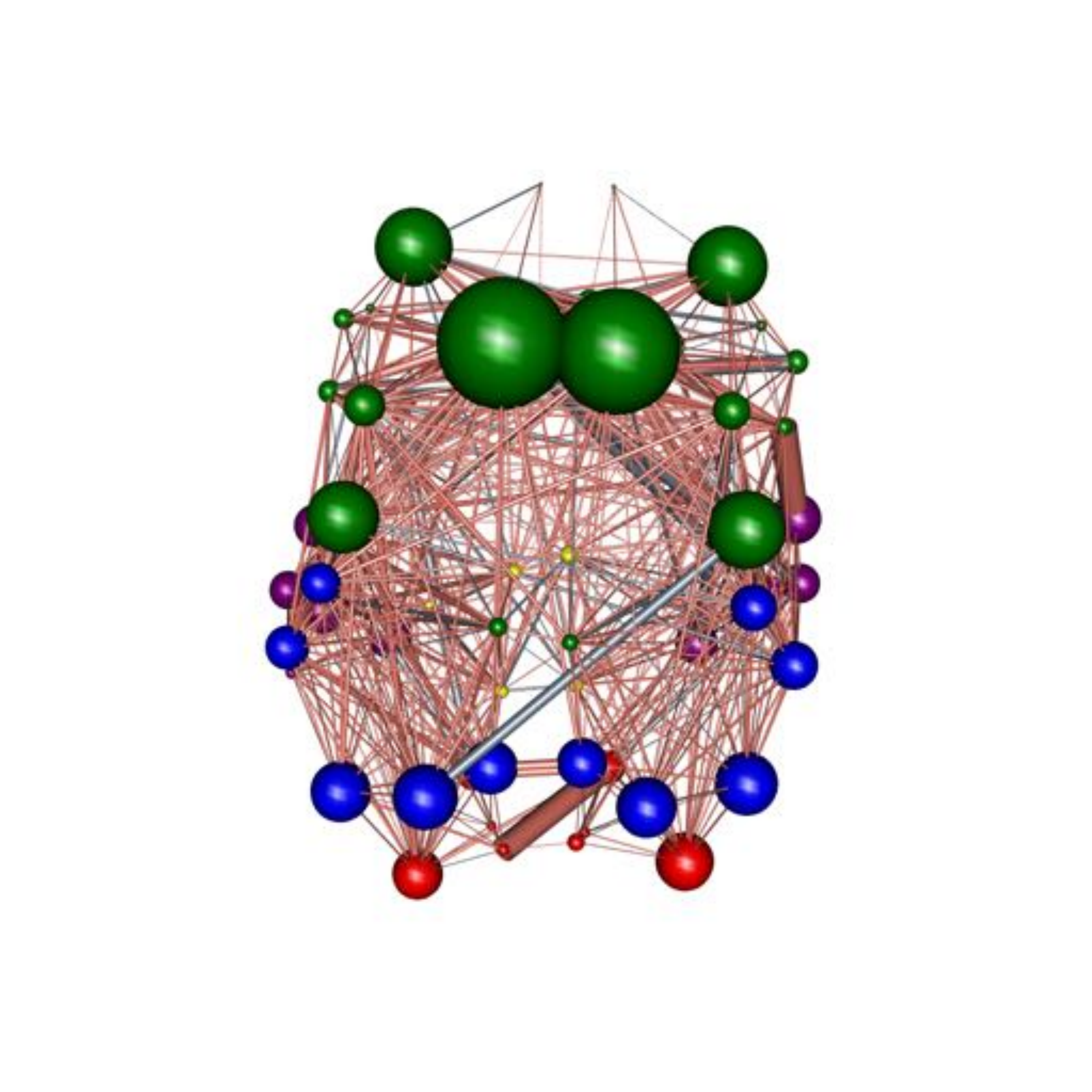}  }  
\parbox{0.2\textwidth}{\centering \includegraphics[trim=6.5cm 6.3cm 6.5cm 5.9cm, clip=true, width=0.2\textwidth]{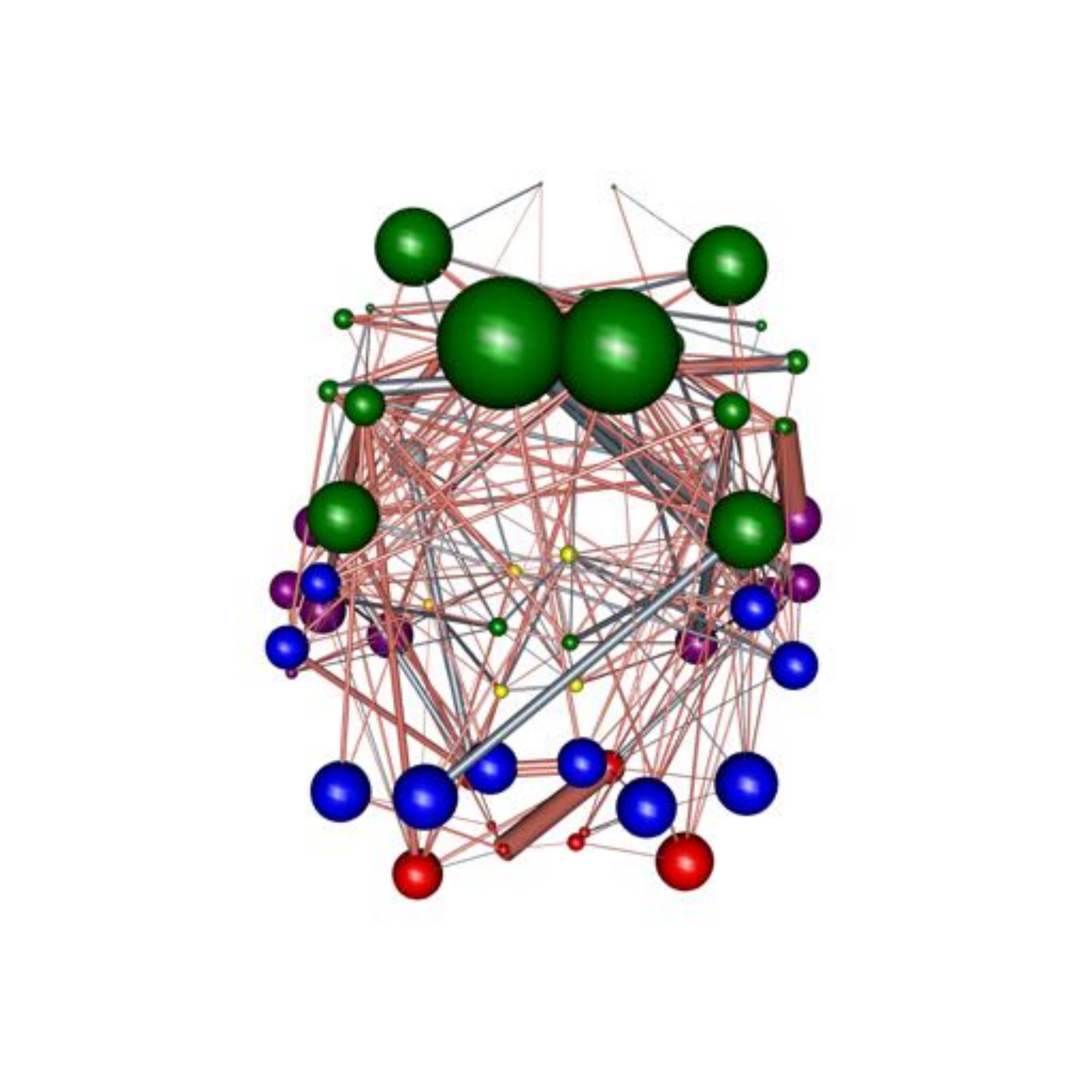}  }  
\parbox{0.2\textwidth}{\centering \includegraphics[trim=6.5cm 6.3cm 6.5cm 5.9cm, clip=true, width=0.2\textwidth]{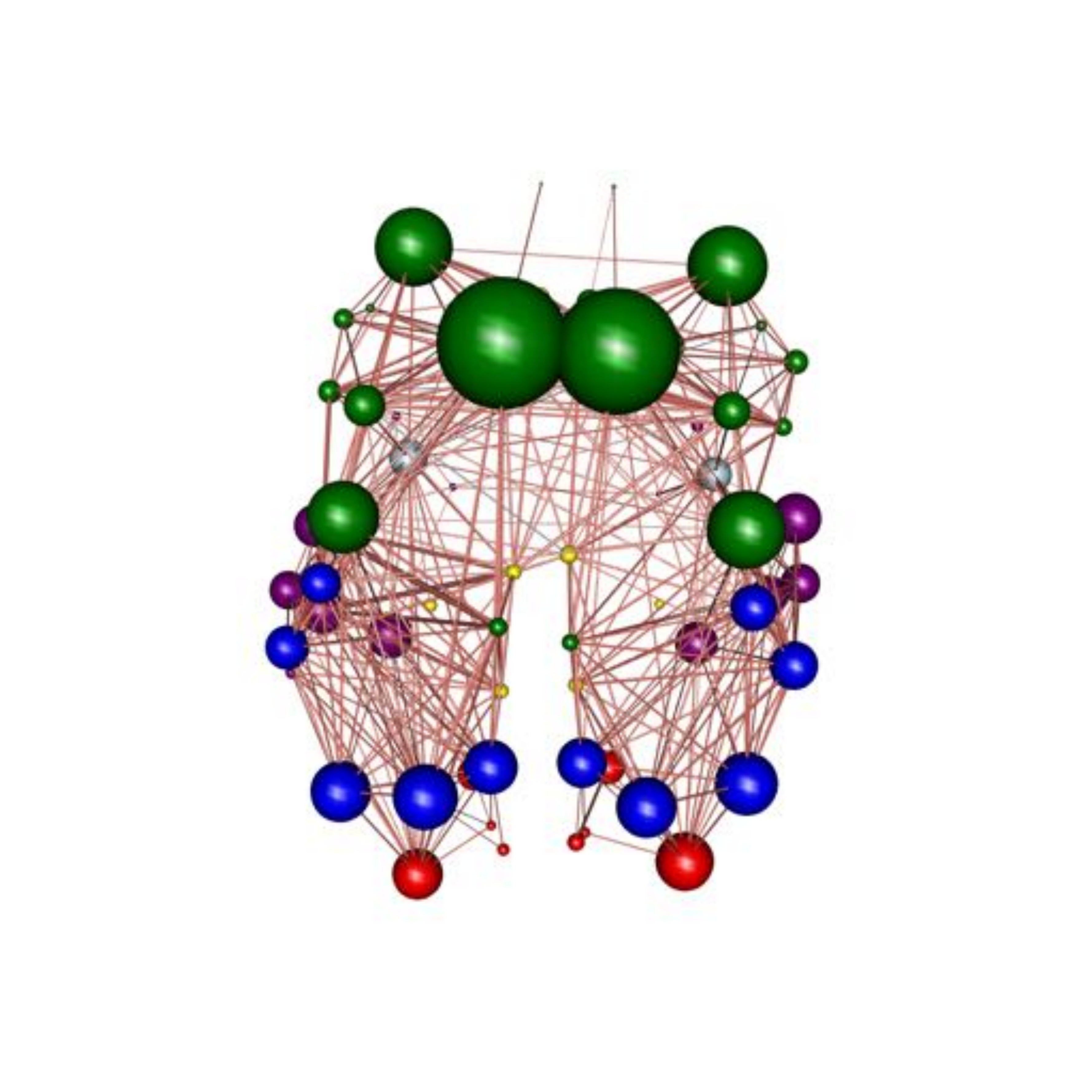}  }  
\parbox{0.2\textwidth}{\centering \includegraphics[trim=6.5cm 6.3cm 6.5cm 5.9cm, clip=true, width=0.2\textwidth]{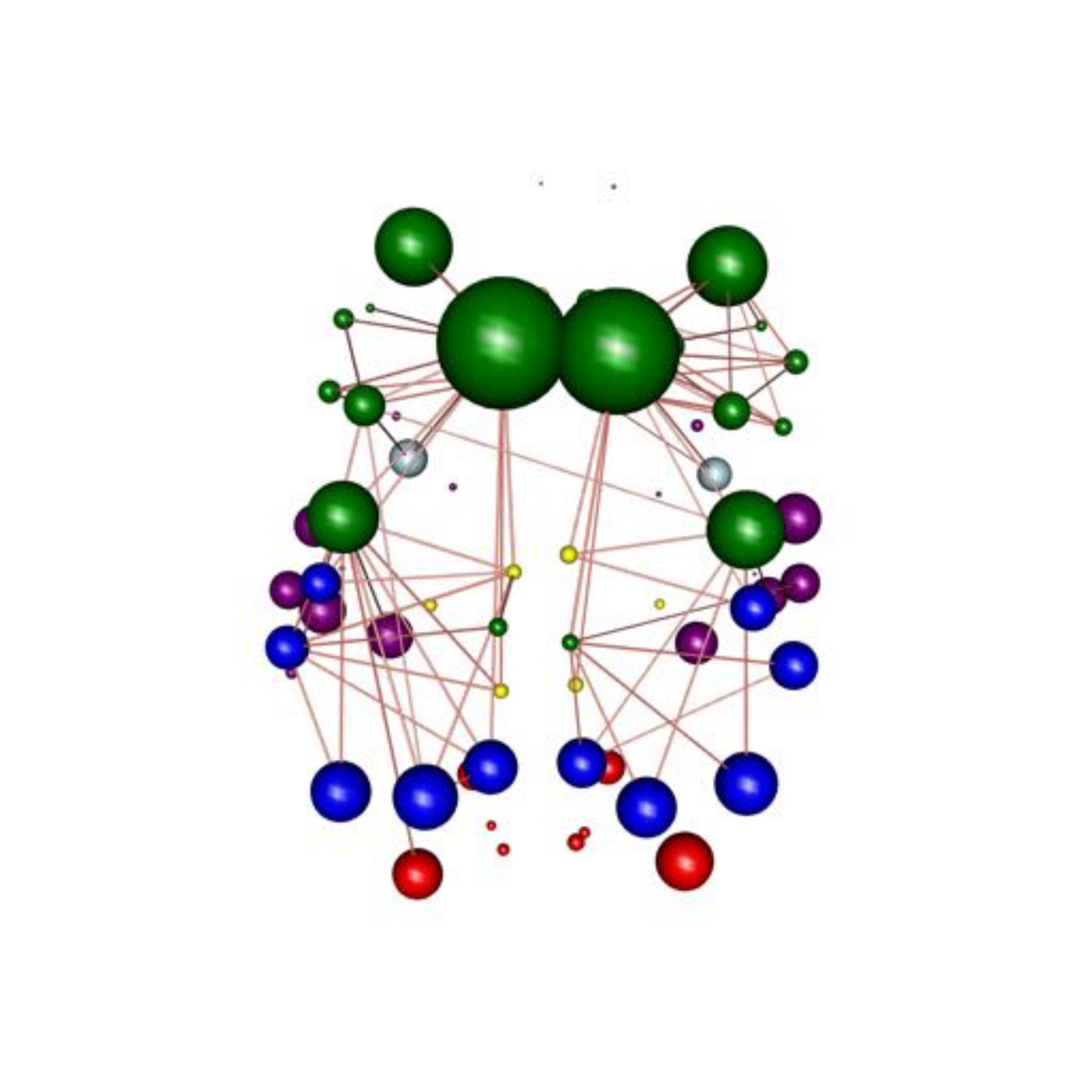}  }  
\parbox{0.05\textwidth}{\centering \includegraphics[width=0.04\textwidth]{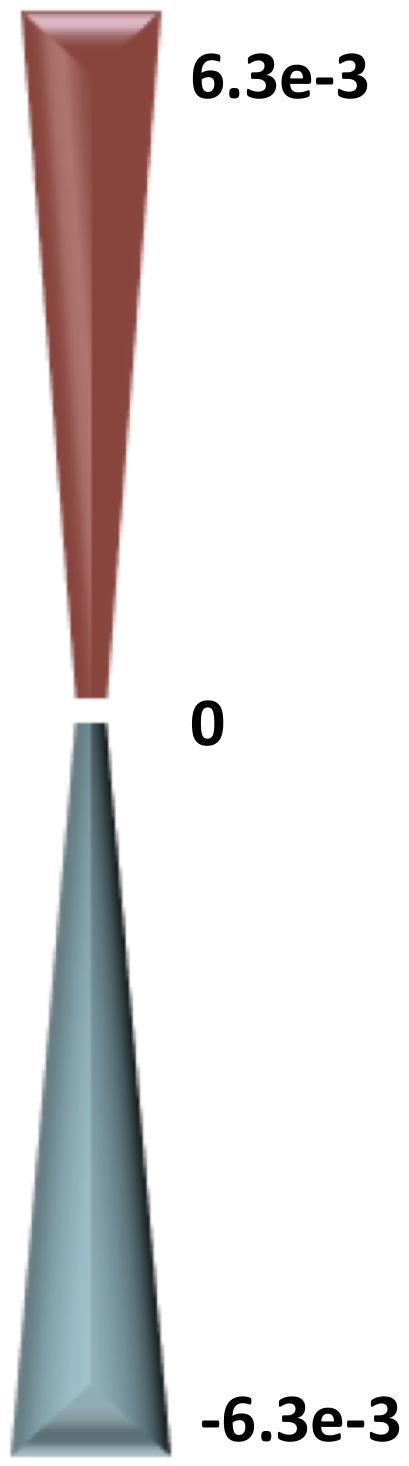}  } 

\centering 
\vspace{1mm}
\parbox{0.01\textwidth}{\centering \rotatebox[origin=c]{90}{\scriptsize{ \large{WODI} }} } 
\parbox{0.2\textwidth}{\centering \includegraphics[trim=6.5cm 6.3cm 6.5cm 5.9cm, clip=true, width=0.2\textwidth]{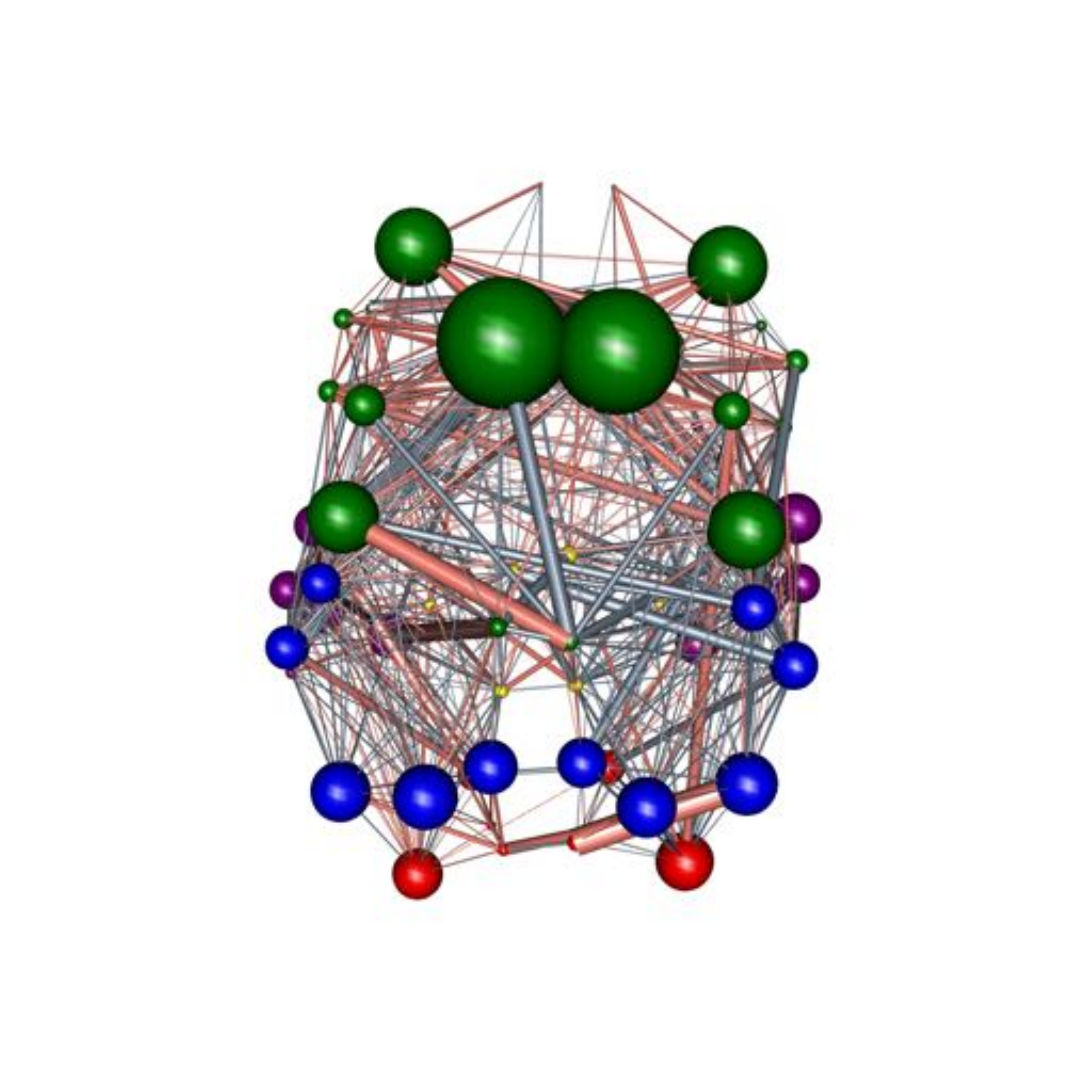}  }  
\parbox{0.2\textwidth}{\centering \includegraphics[trim=6.5cm 6.3cm 6.5cm 5.9cm, clip=true, width=0.2\textwidth]{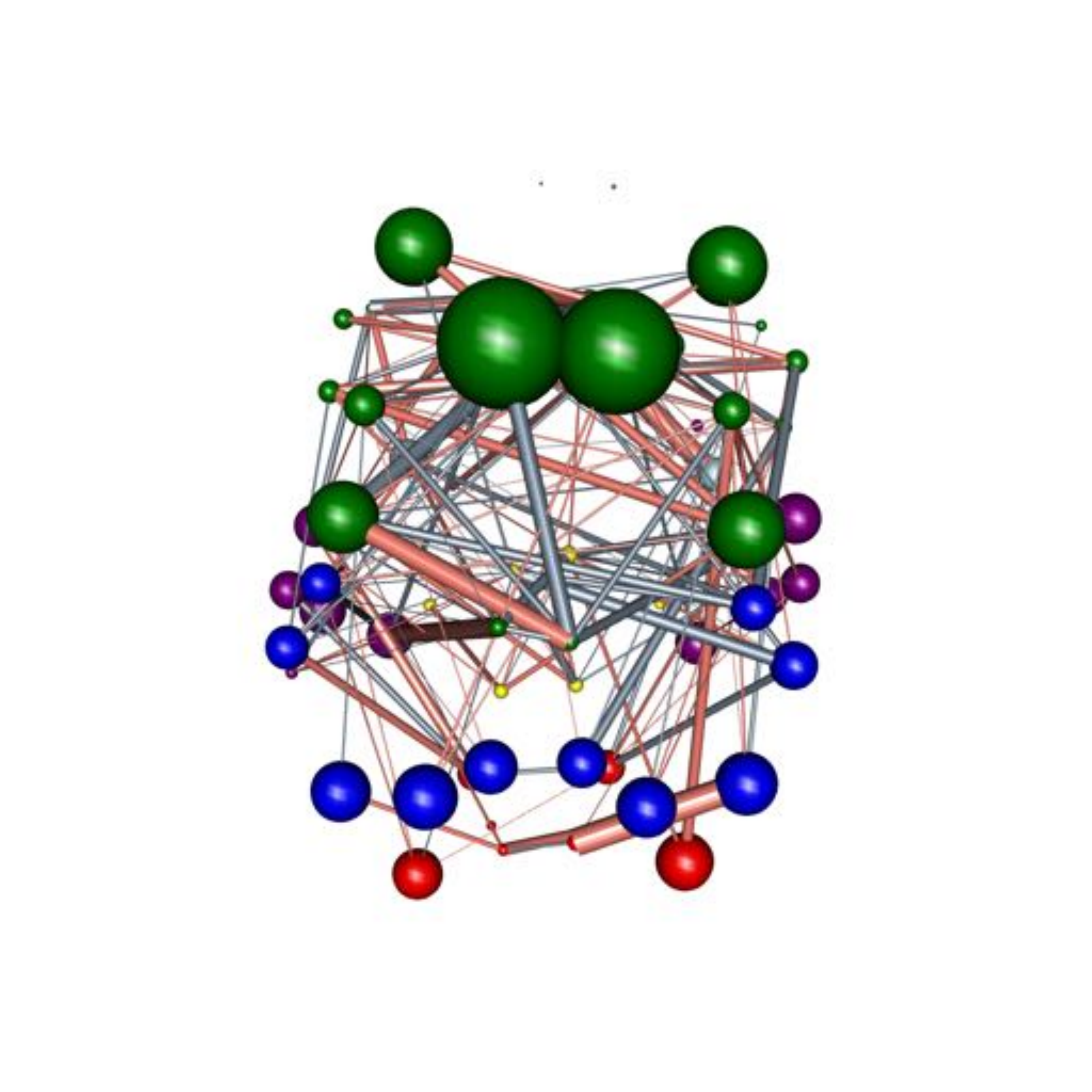}  }  
\parbox{0.2\textwidth}{\centering \includegraphics[trim=6.5cm 6.3cm 6.5cm 5.9cm, clip=true, width=0.2\textwidth]{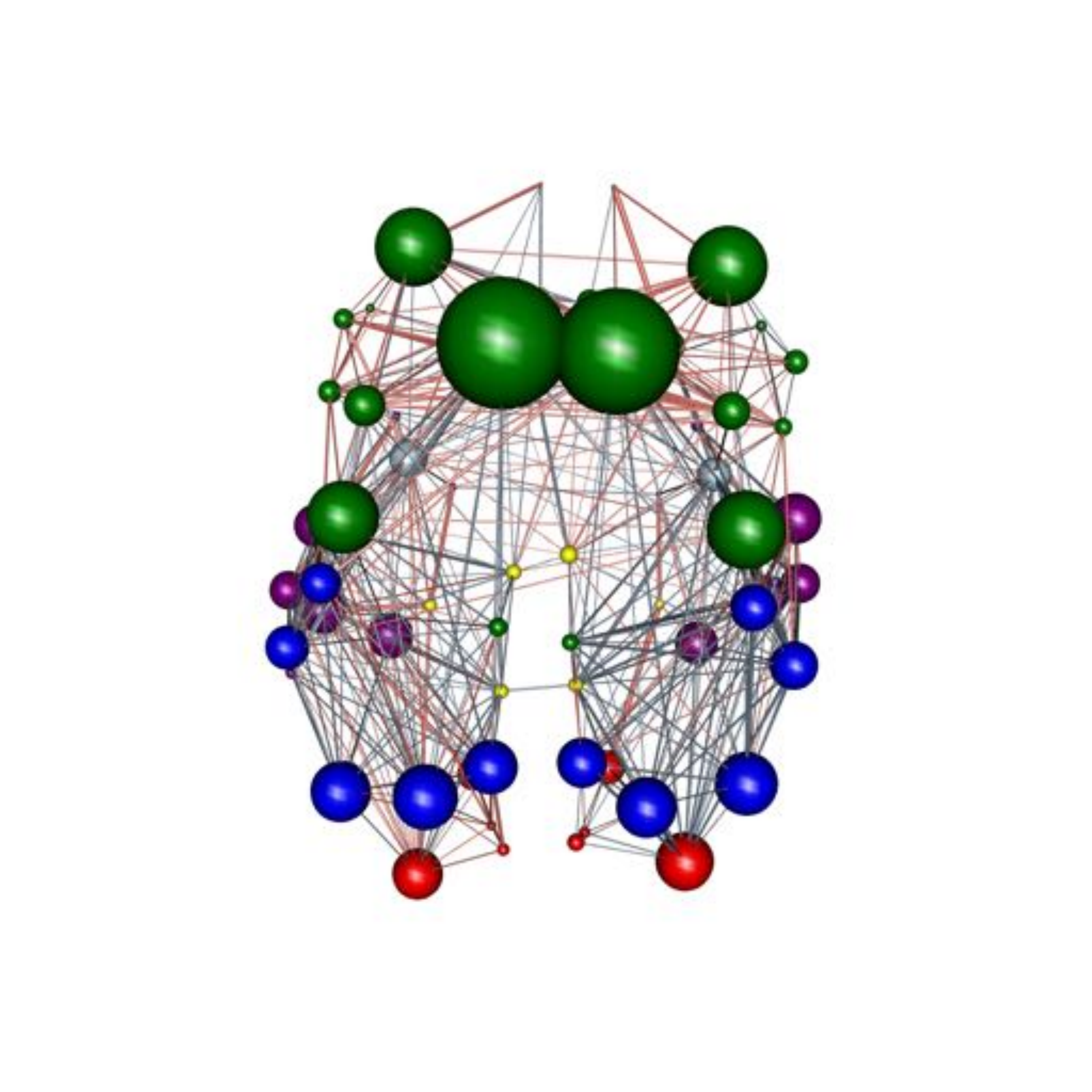}  }  
\parbox{0.2\textwidth}{\centering \includegraphics[trim=6.5cm 6.3cm 6.5cm 5.9cm, clip=true, width=0.2\textwidth]{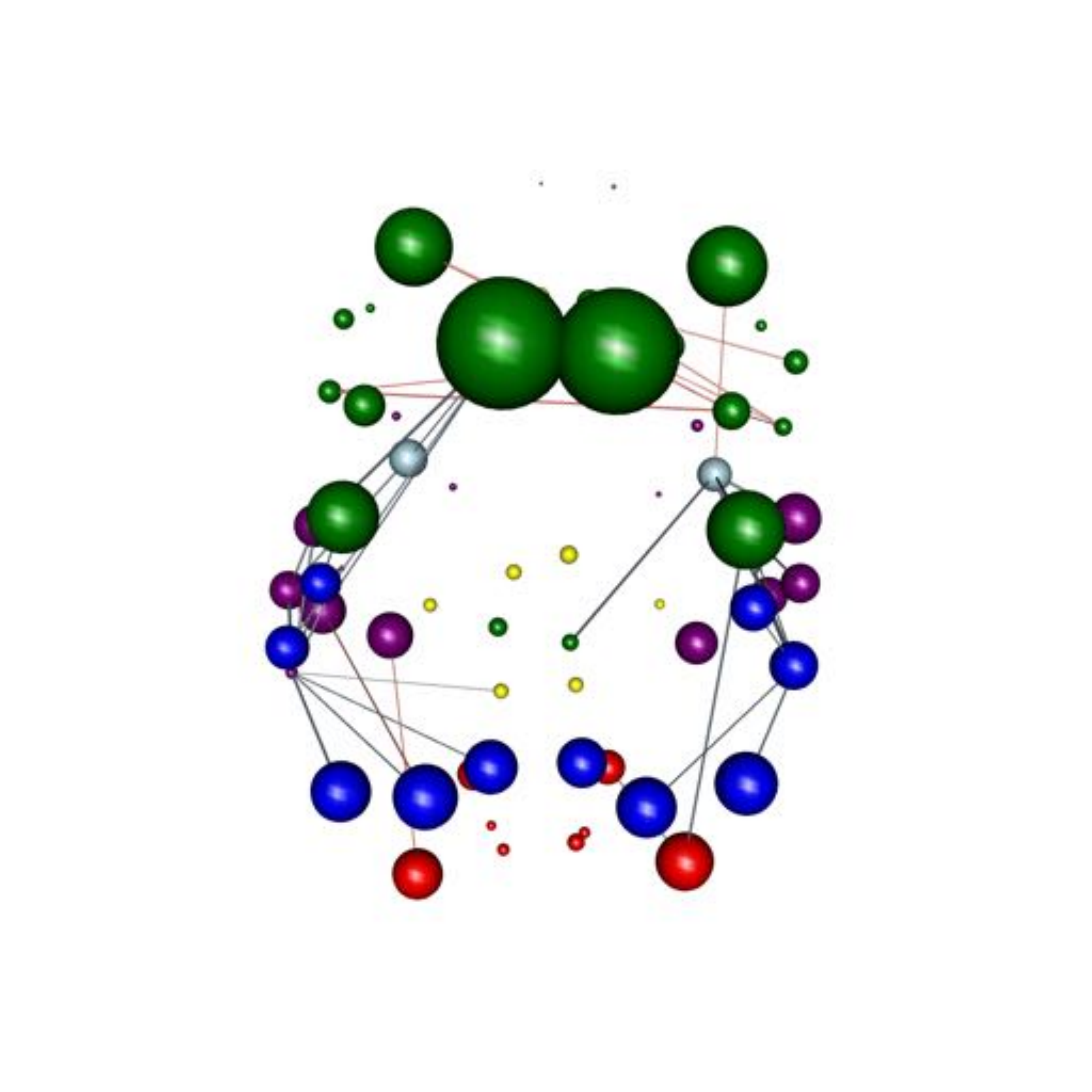}  }  
\parbox{0.05\textwidth}{\centering \includegraphics[width=0.04\textwidth]{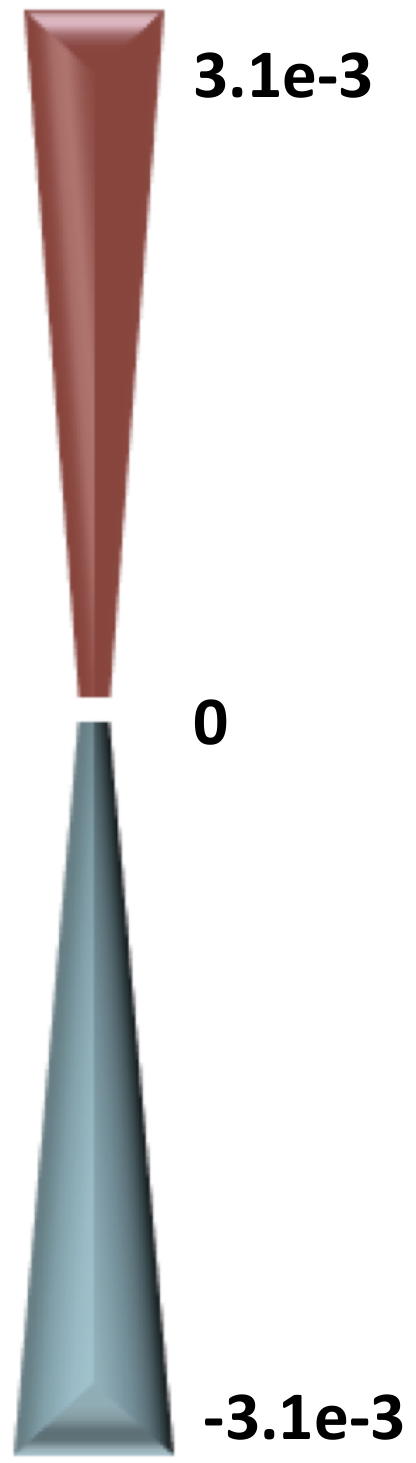}  } 

\centering 
\vspace{1mm}
\parbox{0.01\textwidth}{\centering \rotatebox[origin=c]{90}{\scriptsize{ \large{Wkappa} }} } 
\parbox{0.2\textwidth}{\centering \includegraphics[trim=6.5cm 6.3cm 6.5cm 5.9cm, clip=true, width=0.2\textwidth]{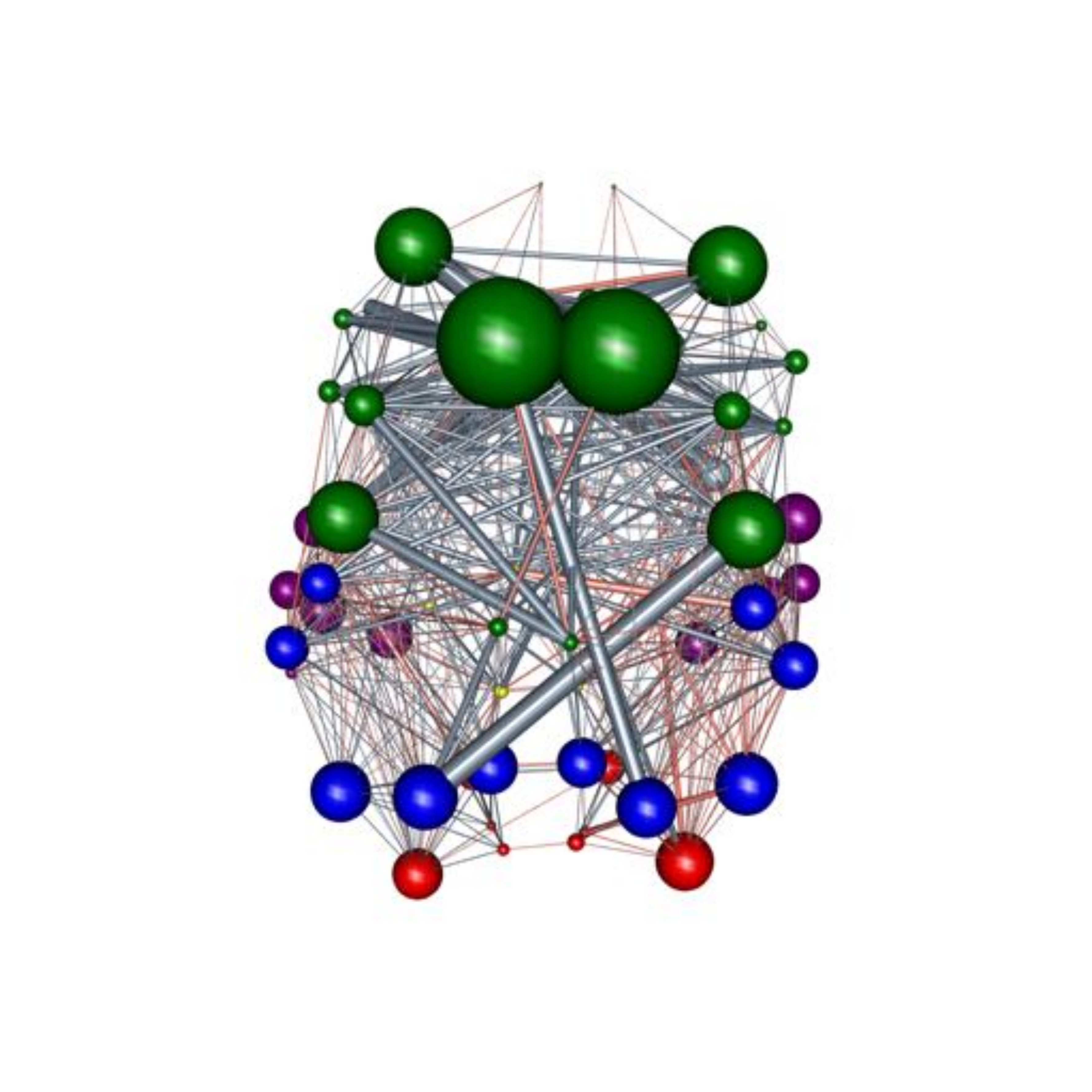}  }  
\parbox{0.2\textwidth}{\centering \includegraphics[trim=6.5cm 6.3cm 6.5cm 5.9cm, clip=true, width=0.2\textwidth]{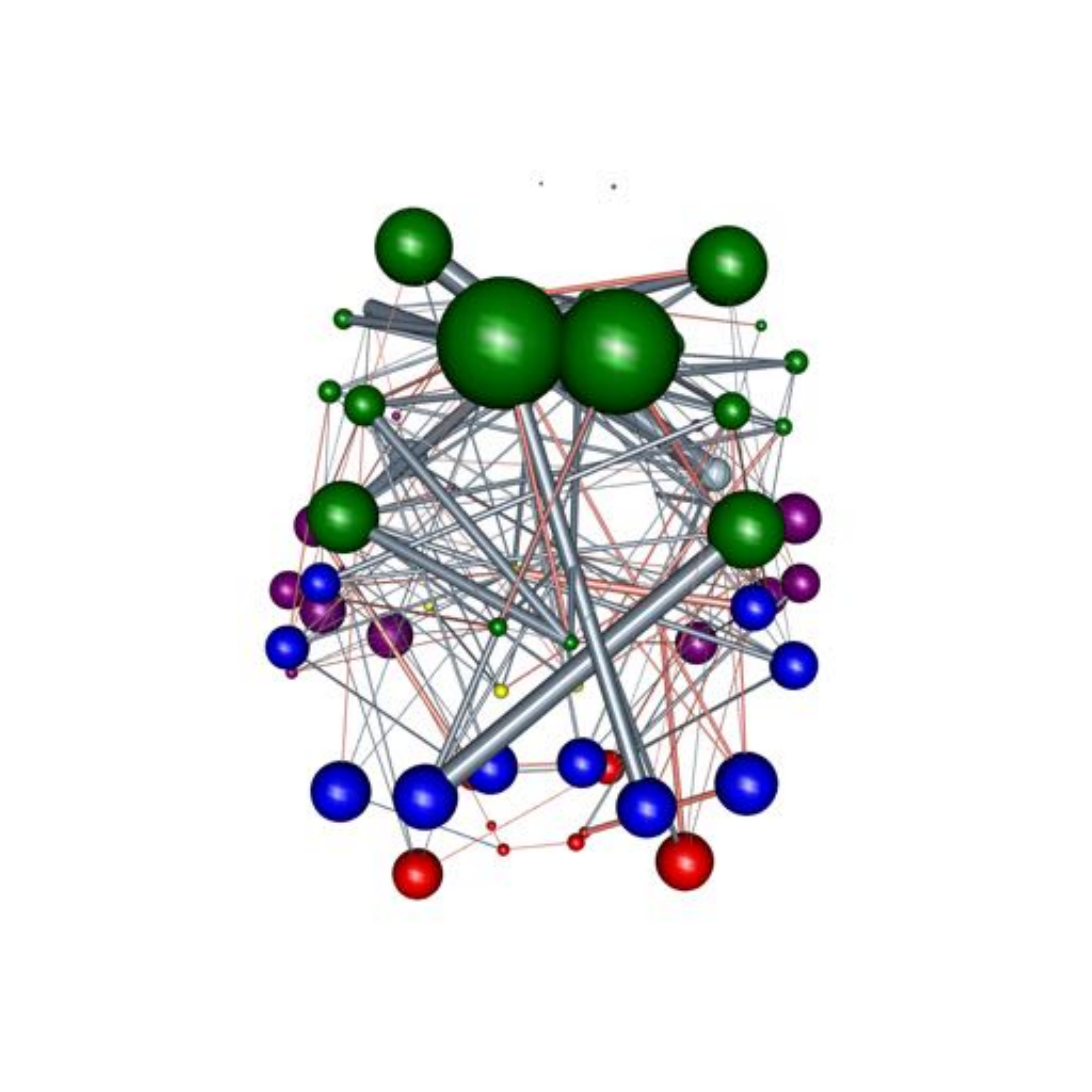}  }  
\parbox{0.2\textwidth}{\centering \includegraphics[trim=6.5cm 6.3cm 6.5cm 5.9cm, clip=true, width=0.2\textwidth]{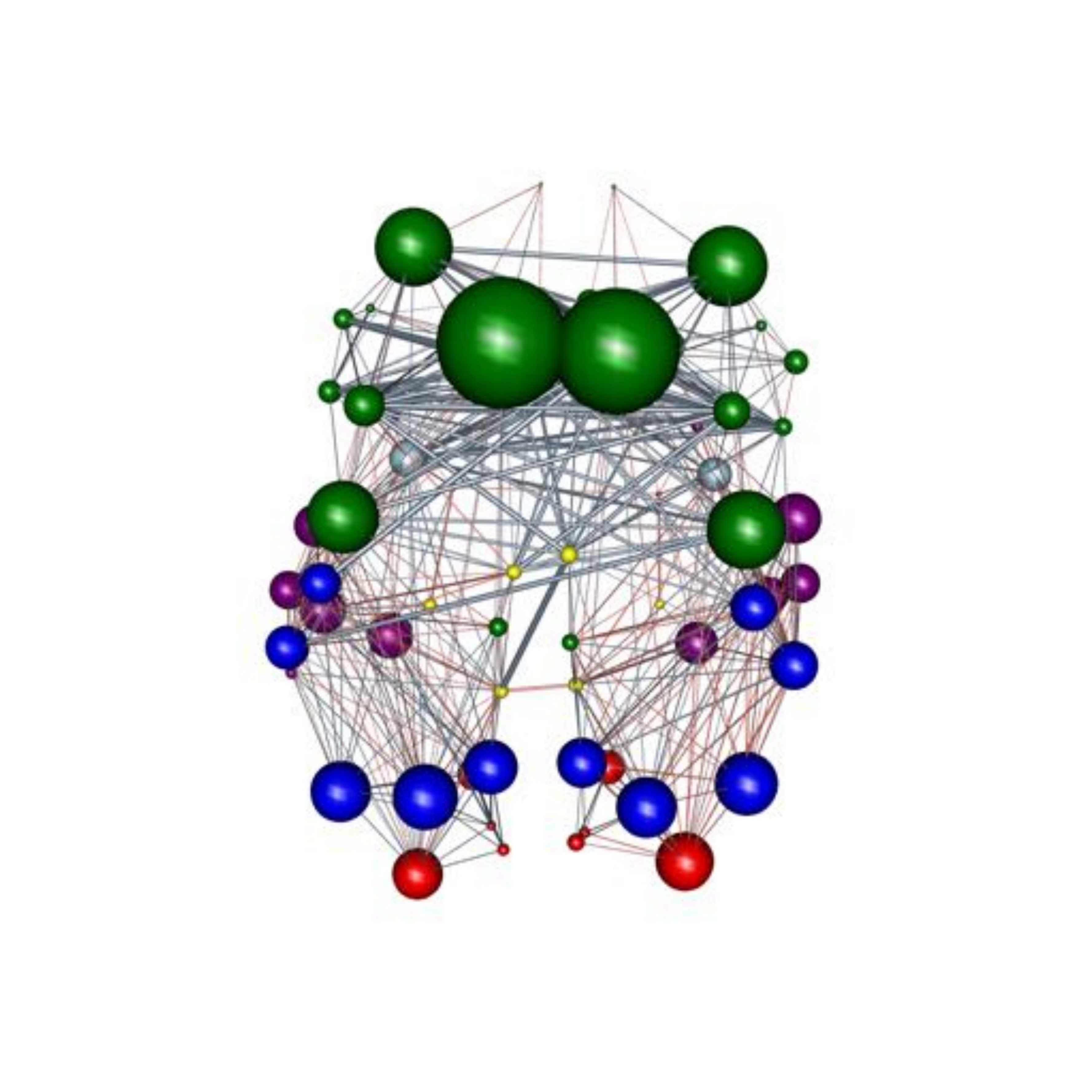}  }  
\parbox{0.2\textwidth}{\centering \includegraphics[trim=6.5cm 6.3cm 6.5cm 5.9cm, clip=true, width=0.2\textwidth]{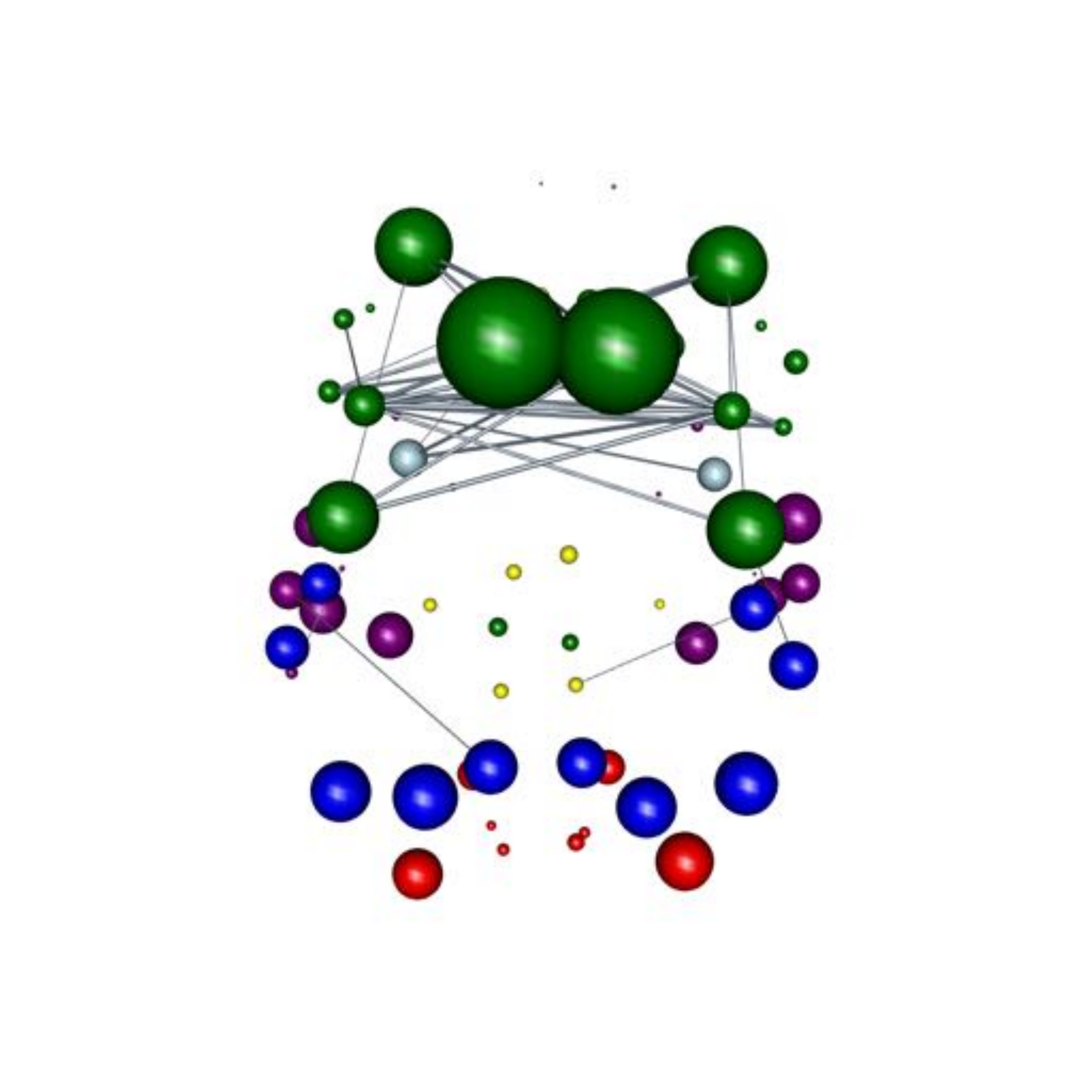}  }  
\parbox{0.05\textwidth}{\centering \includegraphics[width=0.04\textwidth]{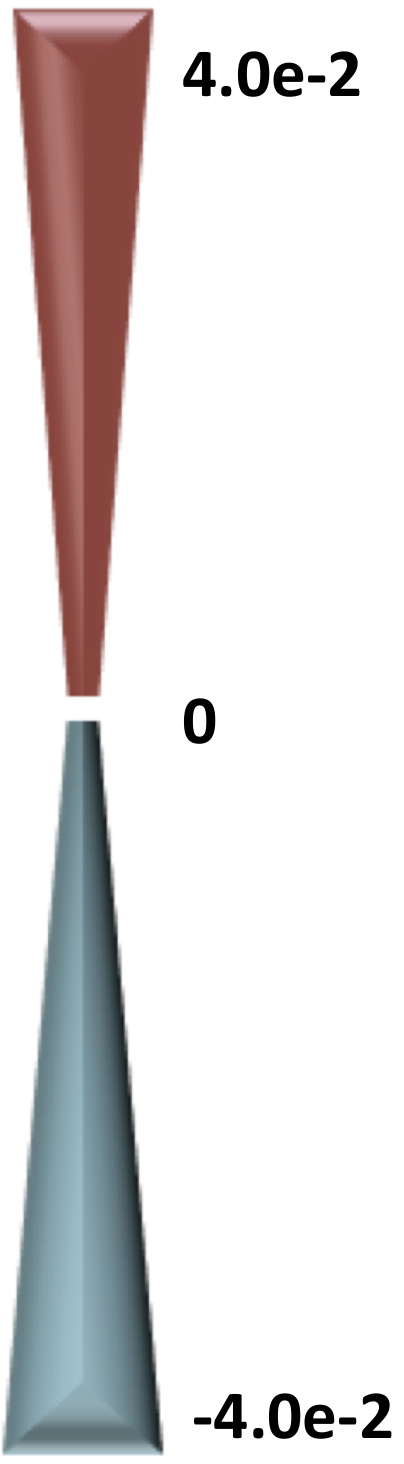}  } 

\centering 
\vspace{1mm}
\parbox{0.01\textwidth}{\centering \rotatebox[origin=c]{90}{\scriptsize{ \large{WISO} }} } 
\parbox{0.2\textwidth}{\centering \includegraphics[trim=6.5cm 6.3cm 6.5cm 5.9cm, clip=true, width=0.2\textwidth]{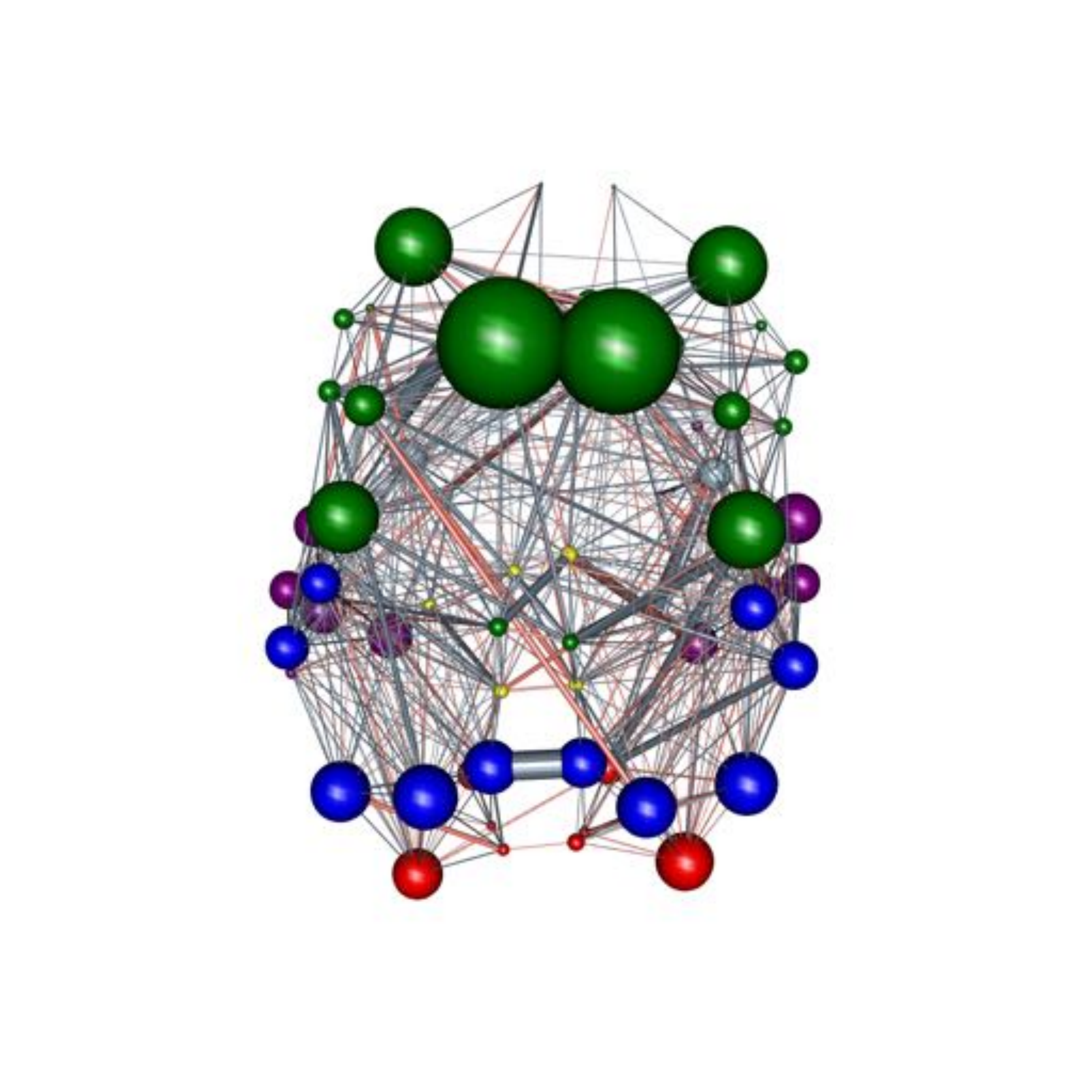}  }  
\parbox{0.2\textwidth}{\centering \includegraphics[trim=6.5cm 6.3cm 6.5cm 5.9cm, clip=true, width=0.2\textwidth]{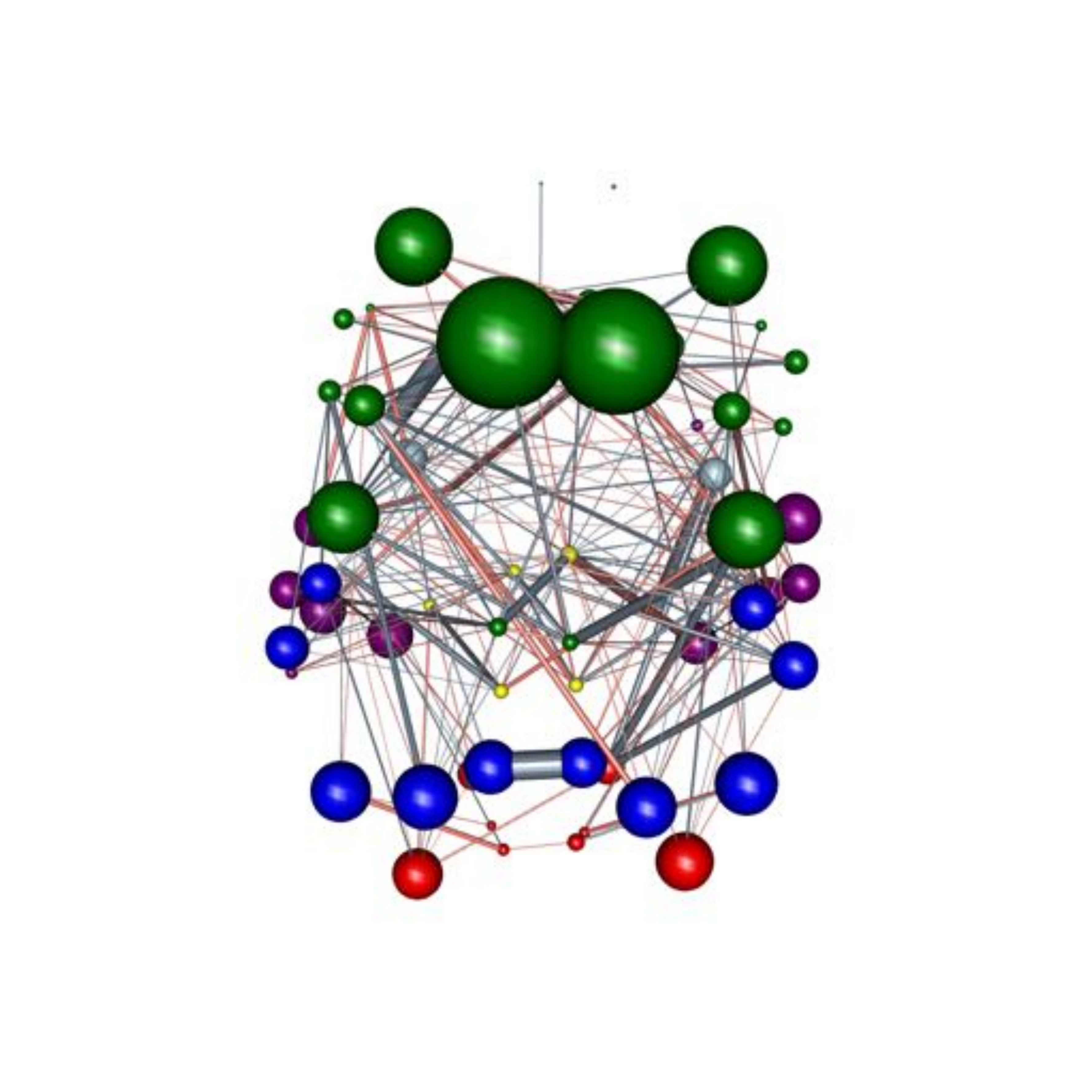}  }  
\parbox{0.2\textwidth}{\centering \includegraphics[trim=6.5cm 6.3cm 6.5cm 5.9cm, clip=true, width=0.2\textwidth]{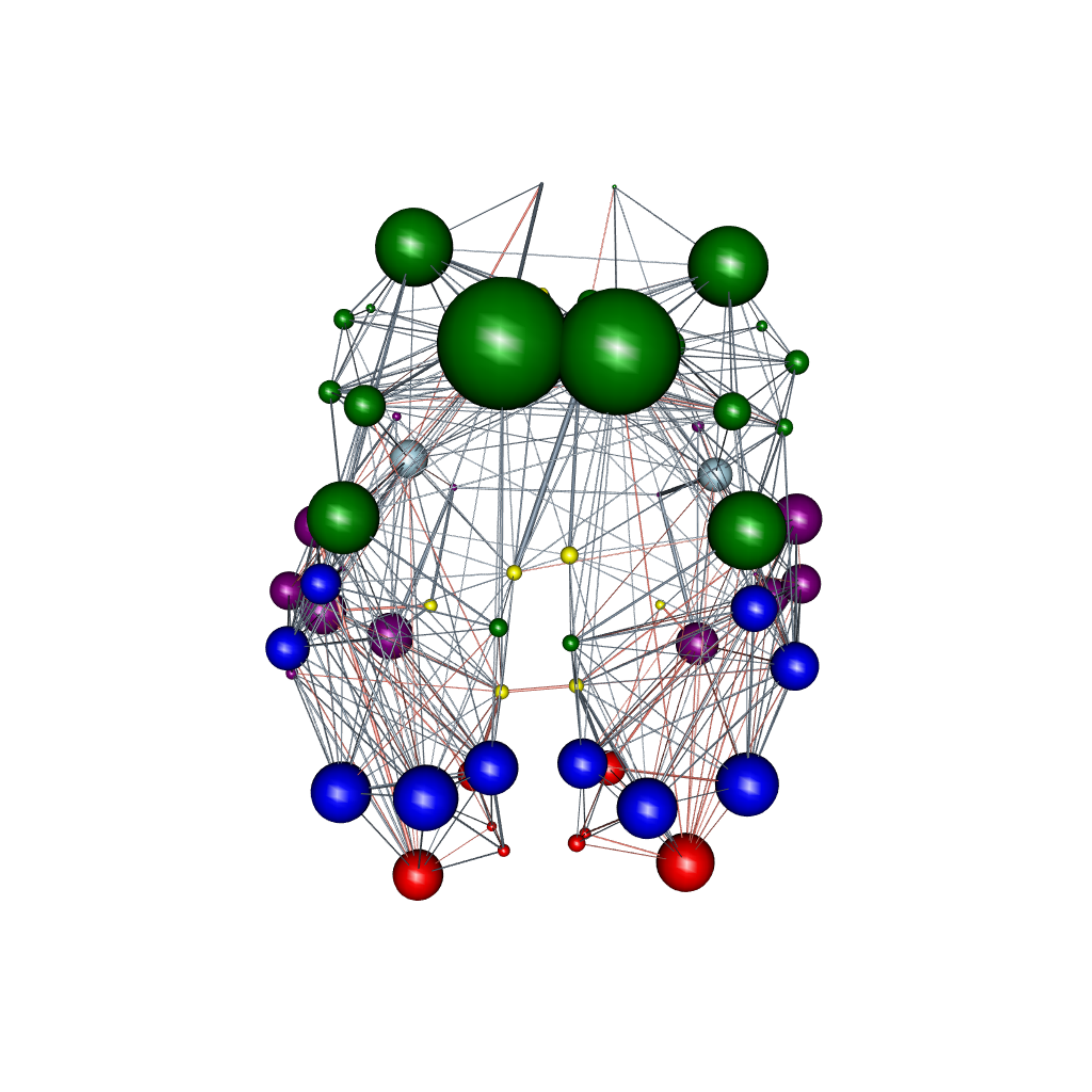}  }  
\parbox{0.2\textwidth}{\centering \includegraphics[trim=6.5cm 6.3cm 6.5cm 5.9cm, clip=true, width=0.2\textwidth]{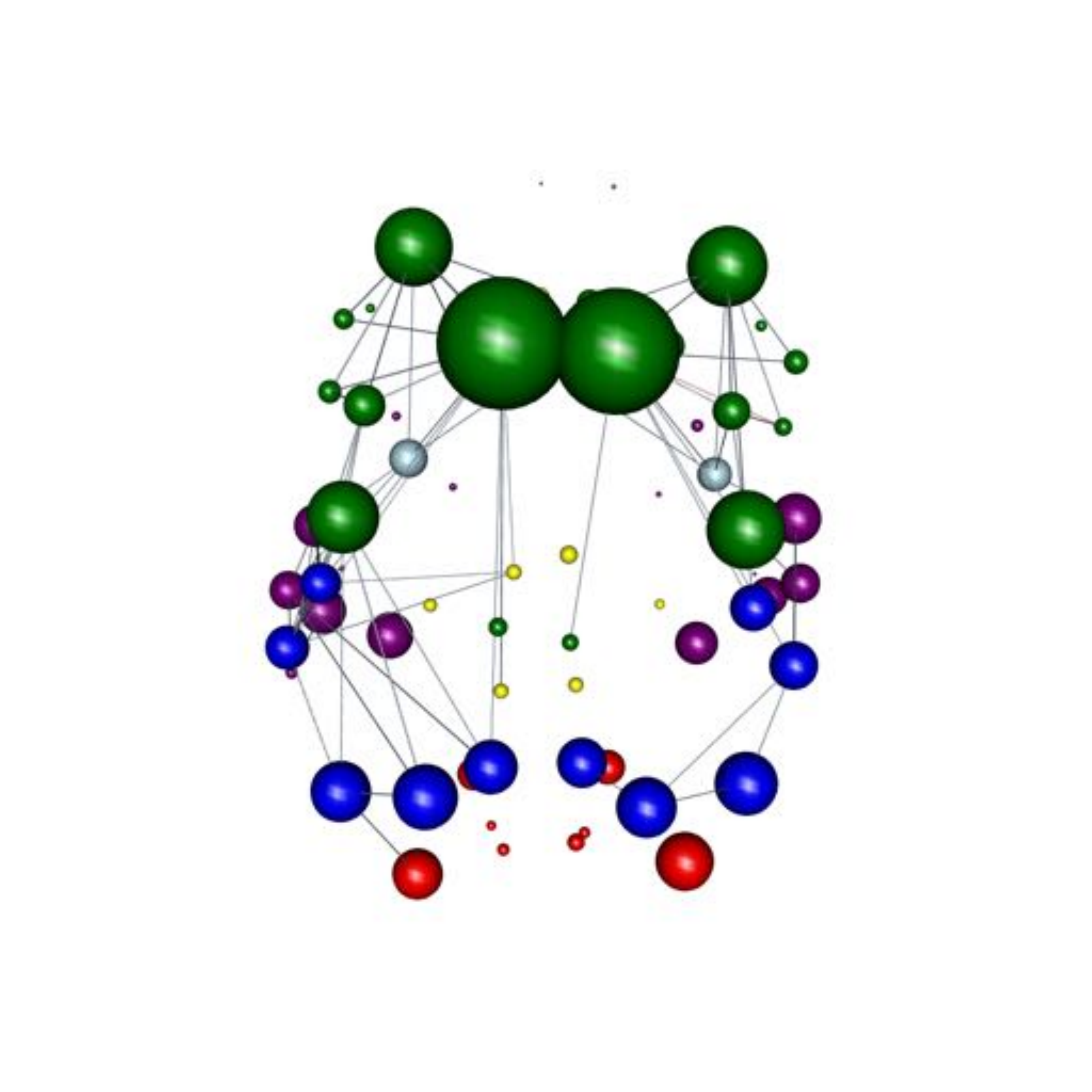}  }  
\parbox{0.05\textwidth}{\centering \includegraphics[width=0.04\textwidth]{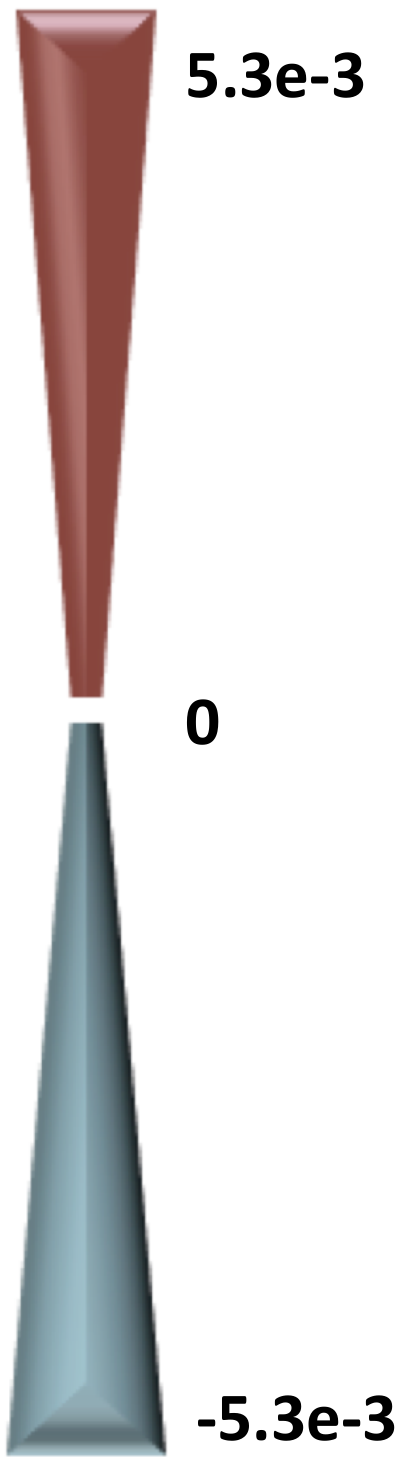}  }

\centering 
\vspace{1mm}
\parbox{0.105\textwidth}{\centering \includegraphics[width=0.02\textwidth]{./frontal.pdf} frontal}  
\parbox{0.105\textwidth}{\centering \includegraphics[width=0.02\textwidth]{./parietal.pdf}   parietal} 
\parbox{0.115\textwidth}{\centering \includegraphics[width=0.02\textwidth]{./occipital.pdf}  occipital}
\parbox{0.125\textwidth}{\centering \includegraphics[width=0.02\textwidth]{./temporal.pdf}  temporal}
\parbox{0.105\textwidth}{\centering \includegraphics[width=0.02\textwidth]{./limbic.pdf} limbic}  
\parbox{0.105\textwidth}{\centering \includegraphics[width=0.02\textwidth]{./insula.pdf}  insula} 

\hspace{0.1mm}

\textbf{\refstepcounter{figure}\label{fig:logmIdentNODDI} Figure \arabic{figure}.}{ Identification results for NODDI based microstructural indices, namely WICVF, WODI, Wkappa, WISO. The first column shows the corrected coefficients for all connections with non-zero values across all subjects. The second column shows the connections that are rejected with a significant p-value ($<0.05$) based on a binomial distribution. The probability of a connection to be selected randomly is defined based on the sparsity of the connectomes and it is equal to 0.04. The third column shows the remaining connections. The forth column shows the connections that are selected significantly above chance (p-value$<0.05$) according to a binomial distribution.}
\end{figure*}

Identification results highlight these relationships between microstructural indices that are are correlated. Thus similar patterns of 'accepted' connections emerge between WF and Wkappa, WF and WODI as well as WMD and WICVF. This is demonstrated in figures \ref{fig:logmIdent}-\ref{fig:logmIdentNODDI}. 
Figure \ref{fig:logmIdent} shows the identification results for tensor-based indices. The first column shows the corrected coefficients for all connections with non-zero values across all subjects. The second column shows the connections that are rejected with a significant p-value ($<0.05$) based on the lower tail of a binomial distribution. The parameters $n$, $p$ of the bionomial distribution are defined based on the number of bootstrap iterations and the probability of a connection to be selected randomly, respectively. The latter is defined based on the sparsity of the connectomes and it is equal to 0.04. The third column shows the remaining connections. The forth column shows the connections that are selected significantly above chance (p-value$<0.05$) according to the upper tail of the binomial distribution with parameters $n$, $p$. Similarly, figure NODDI-based microstructural indices. 
Similarly, figure \ref{fig:logmIdentNODDI} shows the identification results for NODDI-based microstructural indices.

\section{Conclusions}

We propose a new approach to detect both similarities and differences between multi-modal brain networks that it does not depend critically on a threshold. Transportation on a Riemannian manifold is used to constrain the prediction model of functional from structural brain connectivity to SPD and improve performance. The proposed extension of sCCA to SPD sCCA is extrinsic, since sCCA takes place in the tangent space rather than on the SPD manifold. Tangent spaces are only valid in a small neighbourhood and therefore our suggested framework is based on the assumption that the average covariance matrix is close to all subjects' specific functional connectivity matrices. 
Finally, we use the binomial distribution to form a threshold statistic and identify connections that are consistently selected or rejected with probability above chance.

%


\bibliographystyle{IEEEtran}

\bibliography{mriNetworks_final}

\vspace{12pt}

\end{document}